\documentclass[english,aip, numerical,jcp,reprint,twocolumn,floatfix,prd]{revtex4-1}
\usepackage{natbib,hyperref}
\usepackage{amsmath}
\usepackage{graphicx}
\usepackage{hyperref} 
\usepackage{dcolumn}
\usepackage{bm}
\usepackage{epsfig,color,xspace,multirow,xr,bbold}
\usepackage[all]{xy}
\usepackage{setspace}
\usepackage{url}
\usepackage[colorinlistoftodos]{todonotes}
\usepackage{threeparttable} 
\usepackage{natbib,hyperref}
\usepackage{float}
\usepackage{placeins}
\usepackage[utf8]{inputenc}

\usepackage{xcolor}
\newcommand{\RRef}[1]{Ref.~\onlinecite{#1}}
\newcommand{\RRefs}[1]{Refs.~\onlinecite{#1}}
\usepackage{soul} 
\begin{document}
\title{
Revving up $^{13}$C NMR shielding predictions across chemical space: 
Benchmarks for atoms-in-molecules kernel machine learning
with new data for 134 kilo molecules 
}
\date{\today}     

\author{Amit Gupta}
\author{Sabyasachi Chakraborty}
\author{Raghunathan Ramakrishnan}
\email{ramakrishnan@tifrh.res.in}

\affiliation{$^1$Tata Institute of Fundamental Research, Centre for Interdisciplinary Sciences, Hyderabad 500107, India}


\begin{abstract}
\noindent
The requirement for accelerated 
and quantitatively accurate 
screening of nuclear magnetic resonance 
spectra across the small molecules
chemical compound space 
is two-fold: 
(1) a robust `local' machine learning (ML) strategy capturing the effect of neighbourhood on an atom's `near-sighted' property---chemical shielding;
(2) an accurate reference dataset generated with a state-of-the-art 
first principles method for training.
Herein we report the QM9-NMR dataset comprising isotropic shielding of 
 over 0.8 million C atoms in 134k molecules 
of the QM9 dataset 
in gas and five common solvent phases.
Using these data for training, we present benchmark results for the prediction transferability of kernel-ridge regression models with
popular local descriptors. 
Our best model trained on 100k samples, accurately predict isotropic shielding of 50k `hold-out' atoms with a mean error of less than $1.9$ ppm.
For rapid prediction of new query molecules, the models were
trained on geometries from an inexpensive theory. 
Furthermore, by using a $\Delta$-ML strategy, we quench the error below $1.4$ ppm. Finally, we test the transferability on non-trivial benchmark sets that include benchmark molecules comprising 10 to 17 heavy atoms and drugs.


\end{abstract}

\maketitle
\section{Introduction}
Nuclear magnetic resonance (NMR) is an indispensable tool in chemistry, biochemistry and biophysics. It is fast, accurate, information-rich and non-destructive, making it the ideal technique for detecting or describing chemical bonding scenarios. As easy and trivial have most NMR experiments become, it is still a computationally expensive task to estimate NMR shielding tensors or coupling constants for large molecular datasets\cite{helgaker1999ab,mulder2010nmr}. 
While molecules with heavy atoms demand incorporation of relativistic corrections to achieve quantitative accuracy\cite{bagno2003predicting,novotny2016interpreting}, 
computational NMR spectroscopy without such subtle effects are routinely used in
organic chemistry\cite{bifulco2003configurational,cimino2004comparison,seymour2016characterizing,bamine2017understanding,molchanov2018conformational,guzzo2018experimental,sarotti2018structural}. 
For a comprehensive review on computational NMR, refer to \RRef{lodewyk2012computational}. Recently,
Grimme \textit{et al.}\cite{grimme2017fully} discussed the automated prediction of spin-spin coupled $^1$H NMR in various solvents by accessing relevant conformers, to generate experimentally relevant NMR spectra, 
while Buevich \textit{et al.}\cite{buevich2018towards} employed computer-assisted 
structure elucidation algorithms and predicted NMR results to analyse molecular geometries. Lauro \textit{et al.}\cite{lauro2020dft} designed a protocol to identify stereoisomers using experimental and predicted NMR data. 





Amongst many {\it ab initio} quantum chemistry frameworks\cite{Keith1992,Keith1993,Mauri1996,Gregor1999,kutzelnigg1990iglo}, gauge-independent atomic orbital (GIAO) \cite{Ditchfield1972} is the most popular. 
Within the GIAO framework, 
Cartesian components of the
NMR shielding tensor, $\sigma_{ij}^q$,
of a nucleus $q$ 
is calculated as the second-order
magnetic response property\cite{hinchliffe1987ab,gaussmolecular}
\begin{eqnarray}
\sigma_{ij}^q = \frac{\partial^2 E}{\partial B_i \partial \mu_j^q},
\end{eqnarray}
where $E$ is the electronic energy of the molecule,
$B_i$ is a component of the external magnetic field,
and 
$\mu_j^q$ is the $j$-th component of the
magnetic moment of the nucleus $q$. The isotropic
shielding is defined as one-third of shielding tensor's trace, 
$\sigma_{\rm iso}=(\sigma_{11}+\sigma_{22}+\sigma_{33})/3$.
Comparison of predicted values of $\sigma_{\rm iso}$
with experimental results is done by calculating its
`shift' using a standard reference compound
$\delta_{\rm iso}^q=\sigma_{\rm iso}^{\rm ref.}-\sigma_{\rm iso}^q$\cite{mehring2012high}.

$^{1}$H and $^{13}$C are amongst the most commonly studied NMR-active nuclei.
Accurate {\it ab initio} computation of $\delta~^{13}$C requires 
methods such as coupled-cluster singles doubles\cite{price2002computational}, or
spin-component-scaled MP2 with a triple-zeta 
quality basis set to reach a mean error of $<1.5$ ppm\cite{helgaker1999ab,Flaig2014}, albeit incurring a cost which
prohibits the method's applicability for high-throughput studies. 
Composite methods have been proposed---analogous to the G$n$ thermochemistry methods\cite{curtiss2011gn}---that exploit the 
additivity in basis set and correlation
corrections to reach a greater accuracy
\cite{price2002computational,semenov2019calculation}. 
Another method involves tailoring exchange-correlation functionals of Kohn-Sham density functional approximations such as WC04 and WP04\cite{wiitala2006hybrid}.

When relaxing the accuracy 
requirement---while retaining the generality---a density functional approach that has received wide attention, particularly for NMR calculations of both ${}^1$H \& ${}^{13}$C nuclei, is mPW1PW91\cite{adamo1998exchange}. 
This method has been shown to provide good results for acetals\cite{migda2004giao}, pyramidalized alkenes\cite{vazquez2002giao}, acetylenes, allenes, cumulenes\cite{wiberg1999nmr,wiberg2004nmr} and even natural products\cite{bifulco2003configurational,cimino2004comparison}. 
The same approach has also been used to model the
2D-NMR spectrum of exo-2-norbornanecarbamic acid\cite{bassarello2003simulation}. 
Further, a multi-reference standard approach\cite{sarotti2009multi}
has shown consistent estimations of chemical shifts in solutions
with a triple-zeta basis set\cite{sarotti2012application}. 
Thus, even though Flaig \textit{et al.}'s\cite{Flaig2014} benchmark study ranked the B97-2 functional high, next only to the MP2 method, 
the consistency of mPW1PW91
has motivated several works including a recent effort by Gerrard \textit{et al.}\cite{Gerrard2020}, where the authors applied mPW1PW91 with the 6-311G($d$,$p$) basis set to generate NMR chemical shielding and hetero-nuclear coupling constants of molecular components in experimentally characterized organic solids.

While direct application of DFT is feasible for any query molecule,
the questions that arise in chemical compound space (CCS) explorations 
often concern
property trends across large datasets, demanding realistically 
rapid evaluation of the desired property. To this end,
machine learning (ML) based 
statistical inference, in combination with high-throughput {\it ab initio}
computing, offers a viable alternative (see \RRef{cobas2020nmr}). This approach has received
such widespread attention that a
recent competition on the world-wide web,
Kaggle, for 
ML-aided prediction of NMR spectra\cite{bratholm2020community} saw a participation of 2,700 teams across the world.
An earlier proof-of-concept study discussed the feasibility of exploiting the local behavior
of NMR chemical shifts with ML
to achieve transferability to systems that are larger than those used to
train the model\cite{Rupp2015}. That work depended on a cut-off based local 
version of the Coulomb matrix (CM) descriptor\cite{Rupp2012}. Recently, Gao \textit{et al.}\cite{gao2020general} explored deep neural networks (DNNs) in their ``DFT + ML" model and achieved mean-squares error of 2.10 ppm for ${}^{13}$C chemical shifts compared to experimental values. DNN has also been employed for modeling electronic spectra\cite{ghosh2019deep,westermayr2020deep,rankine2020deep}. Kernel-ridge regression (KRR) 
is another ML method offering accuracies comparable to that of DNN
for spectroscopy applications\cite{ML_TDDFTEnrico2015,xue2020machine}. However, 
ML/deep learning may not be limited to single property applications when multiple properties can be explored\cite{ghosh2019deep,westermayr2020neural,pronobis2018capturing}. 

As for descriptors, successive improvements have been made by projecting the three-dimensional  molecular chemical structure into multidimensional tensors\cite{huo2017unified}, four-dimensional hyper-spherical harmonics\cite{bartok2010gaussian}, or a continuous representation such as the variant smooth overlap of atomic positions---SOAP\cite{Bartok2013,De2016}. The latter with Gaussian process regression\cite{Paruzzo2018} predicted chemical shifts with root-mean-squares-error (RMSE) of 0.5/4.3 ppm for $^1$H/$^{13}$C nuclei on 2k molecular solids, while with KRR it was successful in predicting \textsuperscript{29}Si and \textsuperscript{17}O NMR shifts in glassy aluminosilicates across a wide temperature range\cite{chaker2019nmr} comparable to fragment-based estimations\cite{hartman2016benchmark}.

The joint descriptor-kernel formalism of Faber, Christensen, Huang, and Lilienfeld (FCHL) uses an integrated Gaussian kernel function accounting for three-body interactions in atomic environment yielding highly accurate results for
global molecular properties such as atomization energies\cite{Faber2018}. 
Recently, FCHL-based KRR has been applied
to model $^1$H, $^{13}$C shifts and $J$-coupling constants between these two nuclei
for over 75k structures in the CSD\cite{Gerrard2020}.
For a test set, which was not part of training, the same study noted mean absolute errors
(MAE) of 0.23 ppm~/~2.45 ppm~/~0.87 Hz (RMSE: 0.35 ppm~/~3.88 ppm~/~1.39 Hz) for 
$\delta~^{1}$H~/~$\delta~^{13}$C~/~$^{1}J_{\rm CH}$, respectively. 

Here, we present gas and (implicit) solvent phase mPW1PW91/6-311+G(2{\it d},{\it p})-level chemical shielding for all atoms in the QM9 dataset\cite{ramakrishnan2014quantum} comprising 130,831 stable, synthetically feasible small organic molecules with up to 9 heavy atoms C, N, O and F---henceforth denoted the QM9-NMR dataset. We apply KRR-ML using training sets drawn from QM9-NMR, benchmark control settings, and rationalize their influence on the performance of large ML models using up to 100k training examples. 
It has been recently noted\cite{dral2020hierarchical} that the $\Delta$-ML approach\cite{ramakrishnan2015big} facilitates better ML accuracies.
Thus, with converged settings, we provide benchmark learning curves for ML and $\Delta$-ML methods based on three local descriptors---CM, SOAP and FCHL. Finally, we evaluate the transferability of the local ML models---trained only on QM9 molecules---to larger systems in non-trivial benchmark sets that include several drug molecules, a small subset of GDB17\cite{ruddigkeit2012enumeration} with molecules comprising 10 to 17 heavy atoms and linear polycyclic aromatic hydrocarbons (PAHs).


\section{Methodology}
Among popular ML frameworks, KRR is one of the most consistent and accurate\cite{faber2017prediction} framework. In KRR formalism, for a query entity (molecule or atom), $q$, a generic property $p$ from a reference (experiment or theory), 
is estimated as a linear combination of radial basis functions 
(RBFs a.k.a. kernel functions)---each centered at one training entity. 
Values of these RBFs are calculated at $q$, then the distances 
between $q$ and $N$ training molecules defined via their descriptors ${\bf d}$ is given as
\begin{equation}
p^{\rm est}({\bf d}_q) = \sum_{t=1}^N c_t k(|{\bf d}_q - {\bf d}_t |).
\label{eq:krr1}
\end{equation}
The coefficients, $c_t$, one per training datum, are obtained through ridge-regression by minimizing the least-squares prediction error
\begin{eqnarray}
\mathcal{L}   & = & \langle {\bf p}^{\rm ref}-{\bf p}^{\rm est} | {\bf p}^{\rm ref}-{\bf p}^{\rm est} \rangle  + \lambda \langle {\bf c} | {\bf c} \rangle  \nonumber \\
             & = & \langle {\bf p}^{\rm ref}-{\bf K}{\bf c} | {\bf p}^{\rm ref}-{\bf K}{\bf c} \rangle  +  
             \lambda \langle {\bf c} | {\bf c} \rangle  
\label{eq:krr2}
\end{eqnarray}
The size of the kernel matrix is $N \times N$, each element defined in close analogy to the right side of 
Eq.~\ref{eq:krr1}, $ K_{ij} = k(||{\bf d}_i - {\bf d}_j ||) $,
$i$ and $j$ going over $N$ training elements, with $||\cdot||$ denoting
a vector norm. Here, for the choice of CM and SOAP descriptors, we used the Laplacian kernel depending on an $L_1$ norm defined as
$ K_{ij} = \exp ( - |{\bf d}_i - {\bf d}_j | / \omega )$, where $\omega$ defines the length scale of the exponential RBF. 
As shown in \RRef{ramakrishnan2017machine}, optimal solution to Eq.~\ref{eq:krr2} amounts to solving the linear system
\begin{equation}
\left[ {\bf K}+\lambda {\bf I} \right]\ {\bf c} = {\bf p}^{\rm ref.}
\label{eq:krr3}
\end{equation}
The second term in the r.h.s.\ of Eq.~\ref{eq:krr2} is apparent in Eq.~\ref{eq:krr3}; if 
the definition of the descriptor does not differentiate any two training entries, then ${\bf K}$
becomes singular and a unique solution to Eq.~\ref{eq:krr3} can only be found with 
a non-zero value for the regularization strength, $\lambda$. Both $\omega$ and
$\lambda$ constitute hyperparameters in the model, that require a cross-validated optimization
before out-of-sample predictions. Any non-zero value of 
$\lambda$ determined by cross-validation is an indication of the presence of 
redundant training entries, either due to data-duplication or 
 poor quality of the descriptor.
As shown in \RRef{ramakrishnan2015many}, in the absence of redundant training entries,  
$\lambda$ can be set to zero and the learning problem translates 
to solving ${\bf K} {\bf c} = {\bf p}^{\rm ref.}$. Alternatively, 
when linear dependencies may be anticipated---due to numerically
similar descriptor differences---rendering 
an off-diagonal element of ${\bf K}$ to be $\approx1$,
a finite $\lambda=\epsilon$ may be 
used to shift the diagonal elements of ${\bf K}$ away from $1.0$ and the lowest eigenvalue away from $0.0$ thereby aiding Cholesky decomposition.

Prior regularization, ${\bf K}$ is a covariance or dispersion matrix with all of its
off-diagonal elements bound strictly in the closed interval $[0,1]$
with unit diagonal elements.
As per \RRef{ramakrishnan2015many}, 
we can estimate $\omega$ independent of property by restricting $K_{ij}$ corresponding to 
 the largest descriptor difference, $D_{ij}^{\rm max}$, to 0.5, as in 
\begin{equation}
\omega_{\rm opt}^{\rm max} = D_{ij}^{\rm max}/\log(2).
\label{eq:singlewidth}
\end{equation}
In the present study, we also explore the performances of the choices of $\omega$
based on $D_{ij}^{\rm mean}$ and $D_{ij}^{\rm median}$ that will differ from the value of $\omega_{\rm opt}$
based on $D_{ij}^{\rm max}$ depending on the diversity of the training set descriptors
\begin{equation}
\omega_{\rm opt}^{\rm mean} = D_{ij}^{\rm mean}/\log(2); \,
\omega_{\rm opt}^{\rm median} = D_{ij}^{\rm median}/\log(2).
\label{eq:singlewidth1}
\end{equation}
Later we show how these choices are in close agreement with $\omega_{\rm opt}$ values found by a 
scan to minimize the error for a large hold out set. We also discuss how the kernel matrix constructed
with $\omega_{\rm opt}^{\rm median}$  can be applied to model NMR shieldings from gas and different solvent phases. 
\begin{figure}[hbtp!]
    \centering
    \includegraphics[width=8.5cm]{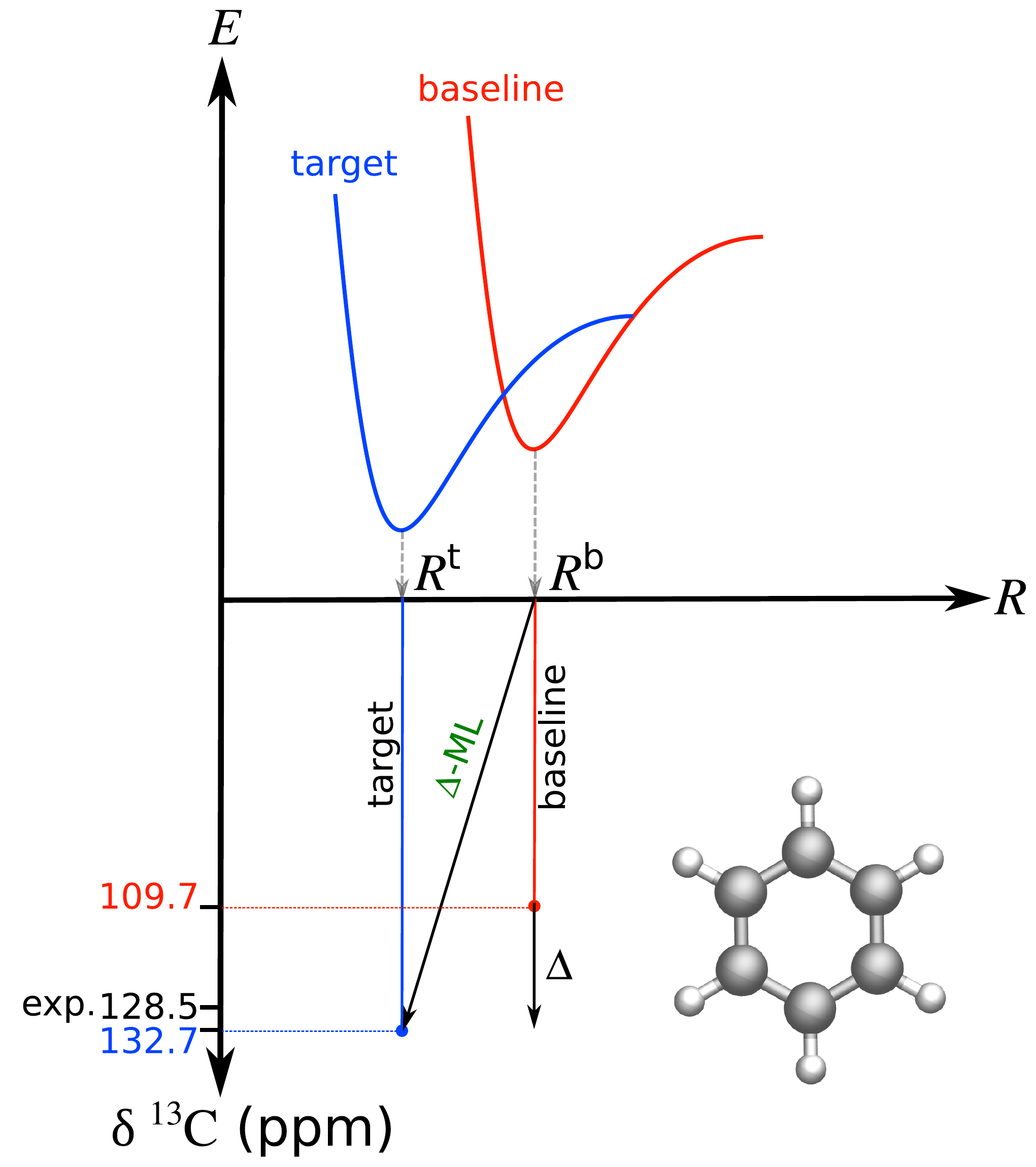}
    \caption{$\Delta$-ML of NMR chemical shifts
    exemplified by benzene; the model is
    trained to predict property from a target-level theory using
    as inputs atomic coordinates and the property from a baseline theory.
    The targetline and baseline shifts of benzene were computed at
    mPW1PW91/6-311+G(2{\it d},{\it p})@B3LYP/6-31G(2{\it df},{\it p}) and
    B3LYP/STO-3G@PM7 levels, respectively.
    }
    \label{fig:nmrscheme}
\end{figure}

The prediction error of an ML model can be unconditionally
quenched with increasing training set size
for a good choice of the descriptor; however, 
the exponential nature of the learning rate often necessitates 
an increase in the model's size by orders of magnitude.  
While the resulting surge in the computational cost associated with the
ML model's execution speed is seldom prohibitive, training with examples of 
the order of $10^6$ places too severe hardware restrictions. When such
hardware limit is reached for training, further drop in an ML model's
error can be attained by training on the deviation of the property from 
inexpensive, yet qualitatively accurate baseline values in a 
$\Delta$-ML fashion\cite{ramakrishnan2015big}.
\begin{equation}
\Delta p({\bf d}_q^{\rm bas.}) = p^{\rm tar.}({\bf d}_q^{\rm tar.}) - p^{\rm bas.}({\bf d}_q^{\rm bas.})
\label{eq:krr4}
\end{equation}
The ML problem now 
involves solving for ${\bf K} {\bf c} = {\bf \Delta p}$. For 
any new prediction, $\Delta$ is augmented with
the baseline
\begin{equation}
p^{\rm est.}({\bf d}_q^{\rm tar.}) = p^{\rm bas.}({\bf d}_q^{\rm bas.}) + 
\sum_{t=1}^N c_t k(|{\bf d}_q^{\rm bas.} - {\bf d}_t^{\rm bas.} |) .
\label{eq:krr5}
\end{equation}
Fig.\ \ref{fig:nmrscheme} illustrates $\Delta$-ML for 
the modeling of NMR shifts with an example molecule.
Often, for any given molecule, 
the determination of minimum energy geometry at the target level incurs a greater computational requirement than that is needed for the estimation of NMR shielding. In the $\Delta$-ML framework, this problem can be alleviated
by using atomic coordinates calculated at the same or a different baseline level for the
construction of descriptors. Hence, new predictions can be rapidly 
made using 
structural information calculated at the baseline-level. 

The formal requirements for a chemical descriptor have been discussed by others\cite{Mauri2017,randic1996molecular,randic1997characterization,Bartok2013,pozdnyakov2020completeness,von2015fourier,Todeschini2000,pozdnyakov2020completeness}. 
Design of structure-based molecular descriptors drew inspiration from the success of generic coordinates such as atom-centered symmetry functions\cite{behler2007generalized,behler2011atom}. Here, we explore CM\cite{Rupp2012}, SOAP\cite{De2016,Bartok2013,engel2019bayesian} and FCHL\cite{Faber2018}. For modeling local properties such as NMR shielding, CM can be truncated by a cutoff radius, $r_{\rm cut}$\cite{Rupp2015}, but there is always the possibility of failing to establish injective mapping between three dimensional molecular structure and the query property\cite{Moussa2012}. Hence, more robust approaches include row-norm sorted CM or ``bag of bonds"\cite{hansen2015machine,huang2016communication,pronobis2018many}. 
Models based on all 3 descriptors show similar prediction times. For a detailed account of solver and prediction times on a single run, see Table.~S2.

\section{Computational Details}\label{sec:comp}
For training data in ML, we 
collected 
B3LYP/6-31G(2{\it df},{\it p})-level
minimum energy geometries of 134k molecules in the QM9 dataset 
 from \RRef{ramakrishnan2014quantum}.
Those structures that have been reported to fragment during the geometry relaxation (3,054 in total) were excluded in this study. 
NMR shielding tensors of selected stable nuclei were calculated at the 
mPW1PW91/6-311+G(2{\it d},{\it p})-level in a single-point fashion
within the GIAO framework\cite{ditchfield1974self,wolinski1990efficient,cheeseman1996comparison} using Gaussian-16 suite of programs\cite{G16short}.
In all DFT calculations, integration grid was
set to \texttt{Ultrafine} with a \texttt{VeryTight} SCF threshold.
To use as a baseline property in $\Delta$-ML, we used 
NMR shielding calculated at the  B3LYP/STO-3G level 
with geometries optimized 
at the PM7 level, the latter done with MOPAC\cite{MOPAC}. 
\textsuperscript{13}C isotropic shielding tensors, $\sigma$, were converted to \textsuperscript{13}C chemical shifts, $\delta$, 
using a reference value for
 $\sigma$ corresponding to that of tetramethylsilane (TMS), which was calculated in gas phase to be 186.97 ppm. 
We have also computed shielding tensors for the entire 131k set 
with implicit modeling of the solvents---carbon tetrachloride (CCl$_4$), tetrahydrofuran (THF), acetone, dimethyl sulfoxide (DMSO), and methanol---with the polarizable continuum model (PCM)\cite{tomasi2005quantum}. This was achieved by invoking \texttt{SCRF} in Gaussian-16 and specifying the solvent name and
retaining default settings.

We have retained the same settings---mPW1PW91/6-311+G(2{\it d},{\it p})@B3LYP/6-31G(2{\it df},{\it p})---to calculate the NMR shielding 
of benchmark molecules that we selected for validating QM9-based ML models.
Initial unrelaxed structures of the linear PAHs studied here have been 
taken from \RRef{chakraborty2019chemical}.
 From the GDB17 dataset, we have randomly selected 8 subsets of molecules, 
each with 25 molecules comprising 10--17 heavy atoms (200 in total). 
Further, we collected drug molecules present in the GDB17 
dataset identified in \RRefs{fink2007virtual,blum2009970,blum2011visualisation,ruddigkeit2012enumeration}. In addition, we also collected
12 somewhat larger drug molecules from \RRef{corey2007molecules}.  
The corresponding SMILES strings of all these `validation' molecules, when available, were converted to initial Cartesian coordinates using the program Openbabel\cite{o2011open}. Initial coordinates of the 12 large drug molecules were created using the Avogadro\cite{hanwell2012avogadro} program. All molecules
have been subjected to preliminary geometry relaxation performed with the force field MMFF94\cite{halgren1996merck}.
We used the default settings in DScribe\cite{dscribe} and QML\cite{QML} 
to calculate the SOAP descriptor and the FCHL kernel matrix, respectively.  
All ML calculations have been performed using codes 
written in Fortran90
with 
interfaces to the SCALAPACK\cite{slug} numerical library. 

\section{Results and Discussions}

\subsection{QM9-NMR dataset}
\begin{figure}
    \centering
    \includegraphics[width=8.5cm]{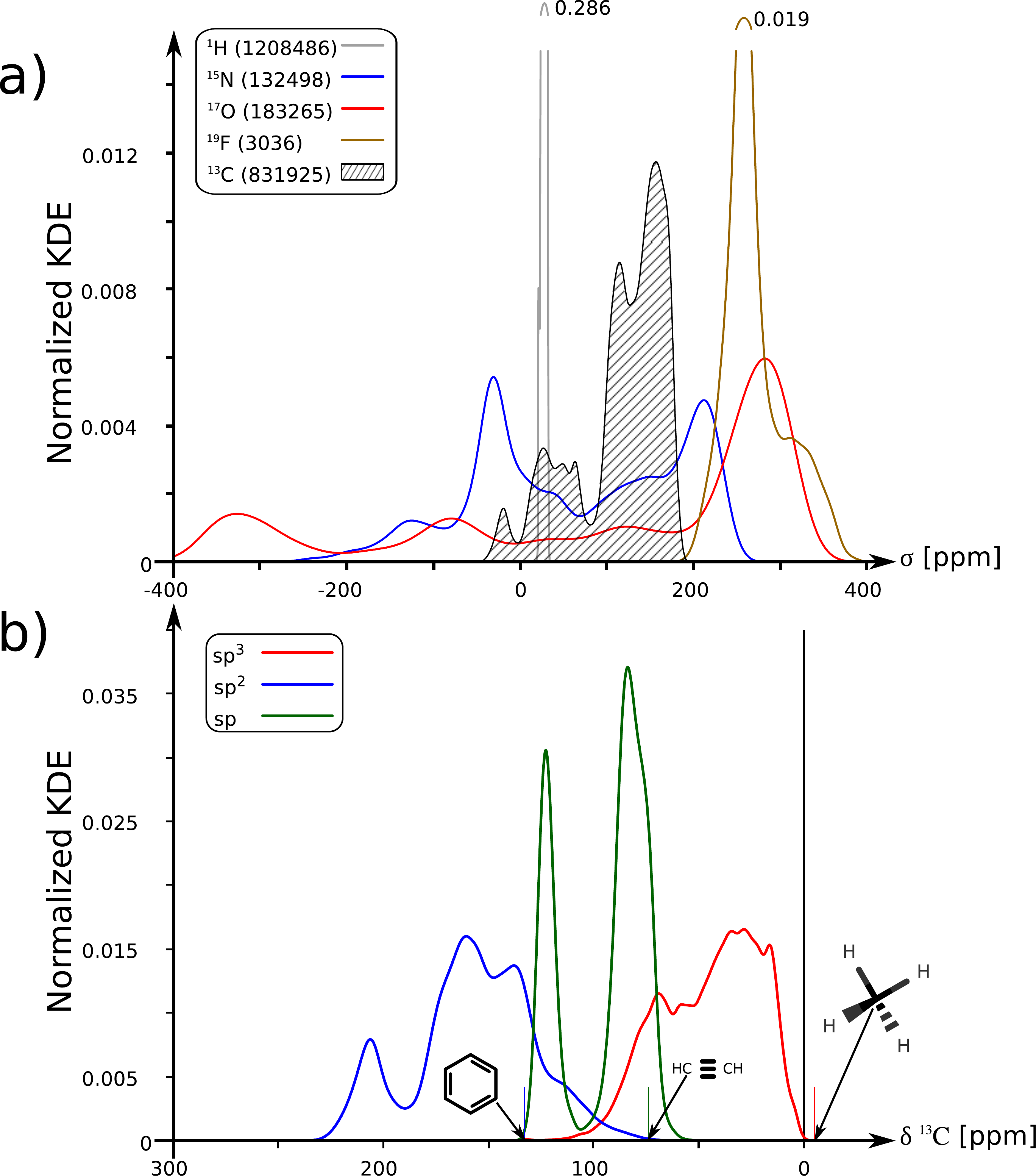}
    \caption{
    Range of NMR shielding properties in the QM9-NMR dataset. 
    a) Element-wise distribution of the shielding ($\sigma$ in ppm) of all the nuclei in 131k QM9 molecules. 
    b) $^{13}$C chemical shifts ($\delta$ in ppm) for the QM9 molecules classified according to hybridization. KDE stands for kernel density estimation.
    }
    \label{fig:KDE_all_elem}
\end{figure}

For a systematic exploration of NMR properties across the QM9 CCS, QM9-NMR dataset was created as per the procedures outlined in Section \ref{sec:comp}. 
This dataset consists of data for stable 130,831 molecules 
amounting to 1,208,486 (1.3 M), 831,925 (832 k), 132,498 (132 k), 
183,265 (183 k), 3,036 (3 k), NMR values for H, C, N, O, and F nuclei,
respectively.
DFT-level NMR shielding of these elements (Fig.~\ref{fig:KDE_all_elem}a) demonstrate the expected range of values. In case of H, the most deshielded nucleus 
corresponds to the one from the cationic ammonium ion 
(encountered in zwitterionic molecules), 
while the most shielded proton belongs to a highly-strained 
secondary amine bonded to N.
Methane offers the most shielded environment for $^{13}$C in QM9, whereas the most deshielded C features in a highly strained multiply-fused-ring molecule. 
Similarly, for N, the most shielded nuclei comes from a strained tertiary amine, whereas for O it features in a strained ring. Most deshielded N and O nuclei belong to a 
zwitterionic molecule.

Besides extrema, Fig.~\ref{fig:KDE_all_elem}a also highlights the chemical diversity of the QM9 dataset. For C and H atoms, majority of the 
NMR shielding parameters come from C({\it sp}$^3$)-H bonds, indicating QM9
to largely comprise saturated organic molecules. 
Unsaturated molecules form a relatively smaller fraction of QM9 as can be seen in its shielding distribution function (between 0-75 ppm for C, Fig.~\ref{fig:KDE_all_elem}b). N atom distribution shows two sharp distribution peaks at about 200 ppm and $-35$ ppm
belonging to primary amine and cyano groups, respectively.
Most frequent
O atoms consist of ether linkages, 
while F atoms show a characteristic broad
distribution around 250 ppm.

NMR shielding values for the entire dataset have also been calculated with
continuum models of five commonly used polar and non-polar organic solvents: acetone, CCl\textsubscript{4}, DMSO, methanol, and THF. 
Formally, a polar solvent will result in a more deshielded environment. However, the influence of the solvent is non-uniform across various C atoms in a molecule depending on the local environment of
an atom in the molecule. Other effects such as hydrogen-bonding, halogen bonding may further influence the chemistry of the molecule resulting in unexpected chemical shifts. Thus, it is necessary to build a database comprising NMR shielding tensors calculated at various solvent media. Here as a first step, we computed the shielding values of the molecules in different solvents under a PCM framework. 
For a better description of the solvent environment it is essential to
go beyond continuum modeling by using micro-solvation models that
account for explicit solute-solvent interactions.
The solvents were chosen
to represent diverse environments:
non-polar, polar aprotic and polar protic. 
For any given $^{13}$C nucleus, the spread in the shielding values
due to the choice of the medium is at the most $\pm$4 ppm (Fig.~S2) with minima and mean at 0 and 0.56 ppm, respectively, suggesting most nuclei to be minimally influenced by the implicit solvent environment. Hence, for ML predictions to differentiate
the results from various phases, it is necessary that the models' prediction
accuracy is much less than $\pm$4 ppm.

 QM9-NMR dataset also contains B3LYP/STO-3G NMR shielding constants for all the 130,831 molecules.
Although the current ML study concerns itself with  NMR shielding of the C-atom, the QM9-NMR dataset can be used to model other nuclei as well. 
To facilitate such and other {\it ab initio} benchmark efforts, 
the entire QM9-NMR dataset, comprising gas and solvent phase results, is now a part of the openly accessible MolDis big data analytics platform\cite{moldis}, {\tt http://moldis.tifrh.res.in:3000/QM9NMR}.



\subsection{AIM-ML modeling of NMR Shielding}
Following the generation of the QM9-NMR dataset with 812k $^{13}$C nuclei, we have selected a random set 
of 100k entries for training the ML models. Further, a separate subset
of 50k nuclei---not overlapping with the 100k training entries---was kept for validating the ML models. 
Additionally, we have compared the distribution of the training and validation subsets
with the total set, and found the normalized density distributions to be 
similar (Fig.~S1). 
Therefore, we believe that the ML models based on large
training sets presented here do not suffer from a selection bias.  

All hyperparameters employed in the ML models have been optimized via
cross validation within the training set. Through a logarithmic grid search, a value of $10^{-3}$ for $\lambda$ was considered appropriate, and was kept constant throughout. NMR shifts are a local property, i.e., for any given atom, only neighboring atoms within a certain `sphere of influence' contribute towards the chemical shift. Hence, 
the effective radius of such sphere, $r_{\rm cut}$, has to be determined empirically. 
Fig.~S3 lists the prediction errors for different descriptors at various cutoff values, in gas and solvent phases. It highlights one clear distinguishing feature between CM/SOAP descriptors, and FCHL; as the cutoff is increased to include more information in the kernel, CM's and SOAP's accuracies show best performances at about 2.3 and 2.0 \AA{} respectively, beyond which the accuracy drops. FCHL, on the other hand, due to the presence of higher-order damping terms, shows a convergent behaviour with the accuracy 
saturating at 4.0\AA{}.

\begin{figure}[t]
    \centering
    \includegraphics[width=8.5cm]{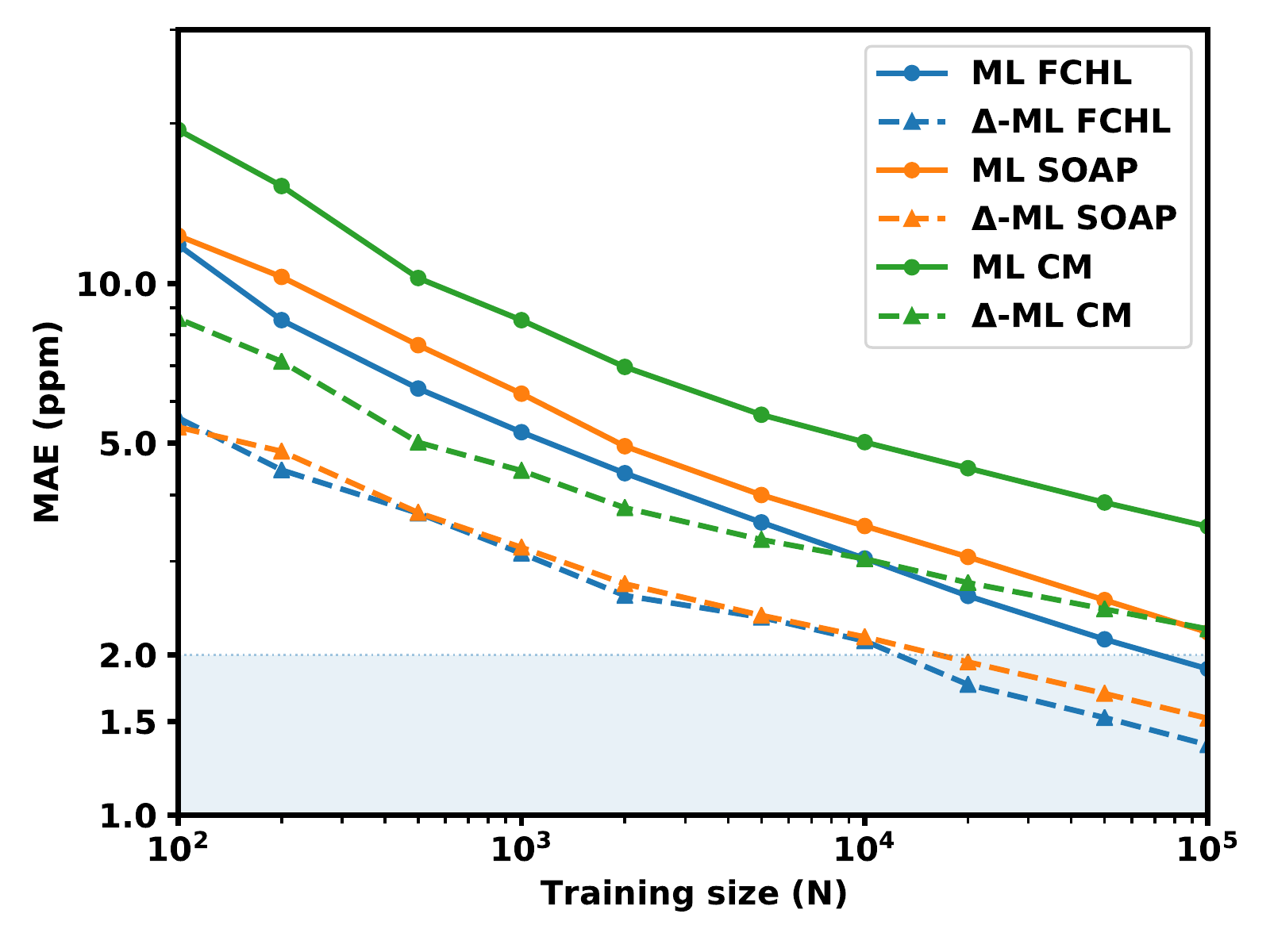}
    \caption{
    ML and $\Delta$-ML out-of-sample prediction errors for CM, SOAP and FCHL descriptors. 
    For increasing trainingset size, $N$, mean absolute error (MAE) in the prediction of NMR shielding of 
    50k hold-out $^{13}$C atoms are shown.
    }
    \label{fig:learning}
\end{figure}

The kernel width, $\omega$, was chosen separately for each descriptor; optimal value of this parameter for a given descriptor is 
inherently coupled to the dimensionality, completeness and the metric of
the ``feature spaces''.
As the numerical values of the descriptor differences need not be same across descriptor definitions (Fig.~S6), a single $\omega$ cannot scale two different kernels with same efficiency.
This fact also sheds light on a heuristic approach for determining  $\omega$---from the descriptor differences, $D_{ij}$,
independent of the property being modeled. We have tested the performance 
of optimal $\omega$ derived using Eq.~\ref{eq:singlewidth1} and found the
values derived using the median of $\lbrace D_{ij} \rbrace$ to perform better than those based on the maximal (Eq.~\ref{eq:singlewidth}) or mean values (Eq.~\ref{eq:singlewidth1}) (Table~S1). For CM and SOAP descriptors, we found these
values to be 422.78 and 18.85, respectively that we have adapted throughout
this study. For these two choices of descriptors, we have also performed
a cross validation to find out the best values 
of $\omega$ 
coinciding with $\omega_{\rm opt}^{\rm median}$ (vertical lines in Fig.~S4). Due to the fact that the FCHL 
implementation in QML does not provide $D_{ij}$ values, a grid search showed the best $\omega$ to be 0.3 (Fig.~S5). Fig.~S7 features the
distribution of the kernel matrix elements based on 10k training examples
for all three descriptors. While the distributions for CM and SOAP
are rather univariate, FCHL's $K_{ij}$ values show a multivariate  
distribution, implying the latter model to be sensitive to the choice of the kernel width. 
After determining the most appropriate hyperparameters for various
choices of descriptors, we collected 10 training sets of sizes: 
100, 200, 500, 1k, 2k, 5k, 10k, 20k, 50k, and 100k. We ensured that each smaller dataset is a subset of a larger one making the learning monotonous. We solved the linear equations of ML (Eq.~\ref{eq:krr3}) using Cholesky decomposition, and the trained machine was used to predict NMR shifts of 50k out-of-sample validation set. 
Mean absolute error (in ppm) for these 50k predictions was admitted as the sole performance metric of the training accuracy (Fig.~\ref{fig:learning}).
It also shows the performances of $\Delta$-ML carried out using B3LYP/STO-3G NMR parameters at PM7 structures.

Overall, one notes from Fig.~\ref{fig:learning} 
that for all descriptors $\Delta$-ML models converging by more than an order of magnitude faster 
(in terms of training set size) than direct ML ones. 
It is also evident that among all three descriptors, FCHL delivers the best performance with an average prediction error of $<2$~ppm; the error drops below 1.4 ppm for $\Delta$-ML modeling. 
However, it may be noted that for training set sizes $\leq$ 10k, both FCHL and SOAP-based
$\Delta$-ML models yielded identical predictions, with SOAP showing an exponential learning rate, while FCHL showing a slightly faster rate going from 10k to 20k training examples---both $\Delta$-ML  models delivering 
$\approx 2$ ppm accuracy already for 20k training. The similarity in performance between SOAP and FCHL can be observed in Fig.~S10 both yielding similar correlation-coefficient across the training set in both ML and $\Delta$-ML. 
For all case-studies, we used FCHL-100k ML and $\Delta$-ML machines.

The origin of accuracy limiting factors in ML was further investigated by categorizing errors based on the NMR shielding range and the representation of the categorized region in the training dataset (Fig.~S9). 
We found that the errors were not uniformly distributed and for certain shielding constant ranges, mean and variance of predictions were more erratic. 
The anomalous error in the -25 to 25 ppm region can be explained by i) under-representation of the aromatic or unsaturated systems in QM9, and ii) larger chemical diversity in the shifts of the unsaturated regions. In the 150--175 ppm region, where we found majority of C shielding values of the  QM9-NMR dataset to lie, 
the prediction errors were rather low and less spread out. We note that
as more data is added in the erroneous regions (such as the -25 to 25 ppm region), the accuracy of the NMR machine improves.

We also probed if the baseline $^{13}$C shielding values computed in 
gas phase can be utilized also for modeling DFT-level values in various
solvents. While it is possible to simultaneously model on
multiple property vectors by feeding in a rectangular matrix---row of column vectors---to the Cholesky procedure, the cost of training can be slightly
minimized by inverting the kernel matrix once and multiplied with any
arbitrary property vector to get new training coefficients\cite{ramakrishnan2015many}.
Table~\ref{tab:solevntphase} demonstrates the versatility of this approach. Inverted FCHL-100k kernel instantly yielded trained machines for all solvents, with sub 2 ppm accuracy. From gas phase to DMSO ($\epsilon=46.8$), we note a modest deterioration in performance. 
\begin{table}[hptb]
    \centering
        \caption{Prediction errors of FCHL-based ML and $\Delta$-ML models, with 100k training examples, in different media. Mean absolute errors
        in the prediction of the NMR shielding of 
    50k hold-out $^{13}$C atoms
        are reported in ppm.}
    \begin{tabular}{l r r}
        \hline
        Medium ($\epsilon$)         &   ML   &     $\Delta$-ML \\
        \hline
        Gas                    &  1.88    &     1.36    \\
        CCl\textsubscript{4} (2.228) &  1.91    &     1.38    \\
        THF  (7.426)                 &  1.99    &     1.48    \\
        Acetone (20.493)             &  1.93    &     1.42    \\
        Methanol (32.613)            &  1.94    &     1.42    \\
        DMSO (46.826)                &  1.99    &     1.49    \\
        \hline
    \end{tabular}
    \label{tab:solevntphase}
\end{table}
\subsection{Application of FCHL-based ML \& $\Delta$-ML models}\label{sec:GDBn}
The magnitude of NMR chemical shift/shielding of a ${}^{13}$C nucleus in a molecule is governed by its local environment. 
The inherent locality of this property implicitly suggests the shielding effect to drop with increasing distance. Subsequently, the information gained from a local moeity of a small test molecule can be reasonably transferred to the same local environment in a large molecule, provided the moeity is not perturbed by chemical interactions alien to the training molecule. 

\begin{figure}[hbt]
    \centering
    \includegraphics[width=8.5cm]{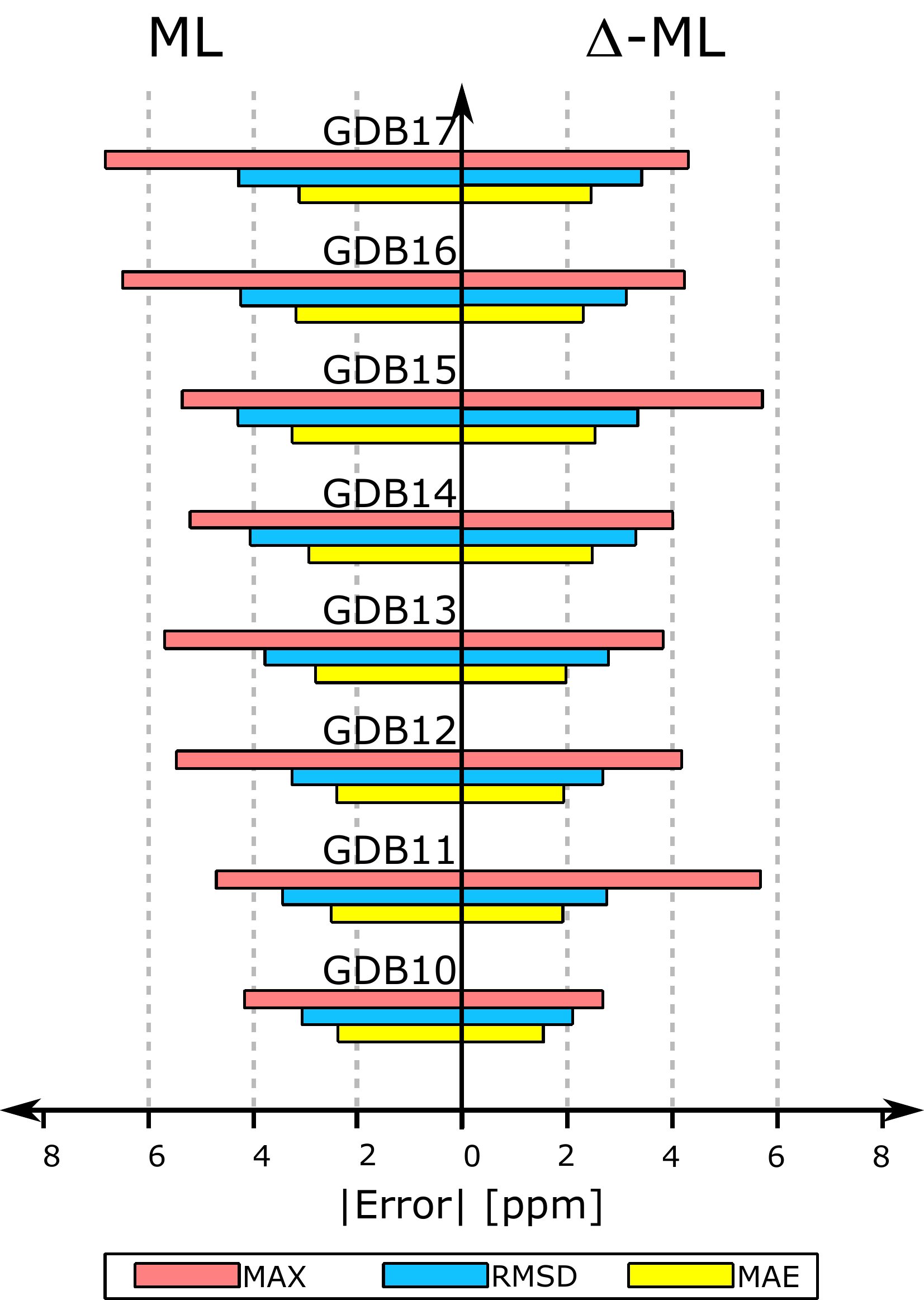}
    \caption{Error metrics for ML and $\Delta$-ML predictions of mPW1PW91/6-311+G(2$d$,$p$)-level $^{13}$C chemical shifts of randomly chosen 25 molecules with $n$ heavy atoms from the GDB$n$ subsets of GDB17; $n=10,\ldots,17$. 
    MAE, RMSD and MAX correspond to mean absolute error, root-mean-squares-deviation and maximal error after averaging over all C atoms in a molecule, respectively.}
    \label{fig:GDB10217}
\end{figure}

Using the 100k ML and $\Delta$-ML machines, we investigate how well these properties can be estimated for larger molecules. 
The graph-based design of the GDB datasets allows one to explore CCS in an unbiased fashion. In Fig.~\ref{fig:GDB10217}, we explore our ML and $\Delta$-ML models' performances across GDB$n$ datasets (where $n=10,11 \ldots, 17$) using 25 randomly chosen molecules per $n$. Each of these 200 molecules were relaxed at the B3LYP/6-31G(2$df$,$p$) level with reference NMR shielding tensors calculated at the mPW1PW91/6-311+G(2$d$,$p$) level (Section~\ref{sec:comp}). 

As expected, $\Delta$-ML generally improves upon ML consistently yielding lower MAE and RMSD. Further, we note maximum average error per molecule (MAX) to overall improve with $\Delta$-ML except for GDB15 and GDB11 possibly due to systems with interactions alien to GDB9. In Fig.~\ref{fig:GDB10217}, ML provides an MAE of $<$ 4 ppm across the datasets while it is usually below 3 ppm for $\Delta$-ML. Arguably, 25 random molecules is not an accurate representation of the entire dataset in question and hence trends instrinsic to these subsets are not transferable across sets. Still, a general observation can be made: increasing number of heavy atoms introduces long-range influences on moieties rendering our machines somewhat inefficient---MAE of both ML and $\Delta$-ML generally increase as we explore larger systems. Thus, apart from small fluctuation in the error trend across GDB10-GDB17---possibly due to sampling bias---the 
local models trained on GDB9 provide quantitative prediction for molecules larger than those used in
training.

GDB$n$ universe contains many drug molecules\cite{fink2007virtual,blum2009970,blum2011visualisation,ruddigkeit2012enumeration}. Fig.~S11 displays 40 such molecules. We tested our models' efficiency in predicting ${}^{13}$C NMR shielding values, and present their error metrics across direct and $\Delta$-ML machines for each molecule. As Spearman coefficients are sensitive to numerical precision, we utilized a modified version by mapping the stick spectrum of the NMR shielding by a step function of height 1 and width of 1 ppm, when needed. 
The largest $\Delta$-ML error encountered in this set is for Desflurane due to the presence of di- and tri-fluoro methyl groups that are under-represented in the training set. The second largest $\Delta$-ML error is for Diethylcarbamazine stemming from deficiencies in baseline data. We note a total of 25 and 5 systems to show MAE higher than 3.0 ppm in ML and $\Delta$-ML modeling, respectively. 
Barring 6 systems, $\Delta$-ML improves upon direct ML's MAE, a trend previously noted in Fig.~\ref{fig:learning} and Fig.~\ref{fig:GDB10217}; not only does it improve MAE, but for 14 systems it also improves $\rho$. Evidently, $\Delta$-ML modeling helps to reach semi-quantitative predictions due to the accuracy of the baseline. In Fig.~S12, we present 12 extra-GDB drugs with their error metrics and DFT chemical shifts. As expected, $\Delta$-ML outperforms direct-ML consistently across all 12 molecules. The highest deviation is noted for Morphine with $\Delta$-ML presenting an MAE $>$ 3.0 ppm possibly because of moeities under-represented in QM9. However, for others the inherent locality of NMR shifts aid prediction. Drugs with extended delocalization present errors, since conjugation is inadequately captured in our ML-models. This deficiency is further noted in Fig.~S13 where predicted chemical shifts of interstitial atoms of PAHs show the maximum deviation.

\section{Conclusion}
We present the QM9-NMR dataset that augments the QM9 set\cite{ramakrishnan2014quantum}---containing DFT-level structures and properties of 134k organic molecules---with NMR shielding values computed at the mPW1PW91/6-311+G(2{\it d},{\it p})-level for about 2.4 million atoms constituting the molecules in this dataset. It may be further extended by including $J$-coupling between $^1$H and other nuclei so that the diverse array of nuclei and properties present in QM9-NMR may aid seamless data-mining or ML studies. The impressive size of the dataset compelled us to explore solvent-phase values using an implicit solvation model, which however may not be adequate to describe effects due to the solute-solvent explicit interactions as addressed in \RRef{molchanov2017solvation}. We focus on predicting the isotropic shielding values of $^{13}$C nuclei in QM9 entries through KRR-ML models with Laplacian kernels. Upon benchmarking the performance of ML models across 3 descriptors: CM, SOAP and FCHL, we note a monotonous improvement in learning with increasing training set size up to 100k, with respect to predictions for a 50k hold-out set, where an FCHL-based (without alchemical corrections) ML-model showed the least MAE of 1.88 ppm. $\Delta$-ML, using PM7 geometries and B3LYP/STO-3G baseline values, improves upon this accuracy to yield an MAE of 1.36 ppm. This is an improvement over the current record in out-of-sample prediction error in data-driven $^{13}$C nuclei NMR shielding modeling\cite{Gerrard2020}. SOAP-based ML model's under-performance could be speculated to the use of Laplacian kernel-based KRR when Gaussian process regression is more effective. The performance drops with increasing diversity of validation molecules but the target being of local nature benefits from our models and aids in the prediction of $^{13}$C shielding in molecules much larger than those in training set. Such a trend has been noted during the validation of $^{13}$C shielding for a random subset of 25 molecules collected from GDB10 to GDB17 sets. Although, the prediction accuracy decreased with increasing molecular sizes, the MAE reported across datasets remained within 4.0 ppm for ML and 3.0 ppm for $\Delta$-ML. When predicting $^{13}$C shielding of drug molecules--- one containing 40 drug molecules from GDB17 universe (Fig.~S11), and the other containing 12 drugs with 17 or more heavy atoms (Fig.~S12)---$\Delta$-ML improves upon ML's performance with the MAE decreasing from 3.7/4.2 ppm to 2.3/2.6 ppm for 40-drug/12-drug datasets, respectively. However, delocalization in linear PAHs (Fig.~S13) proves challenging because of the small cutoff values decided from cross validation on molecules lacking such effects.

While the deficiency in our models should not fade with other local descriptors\cite{langer2020representations}, augmenting the training set with systems displaying extended conjugation such as PAHs, fullerenes etc., or improving upon the current baseline for $\Delta$-ML should lead to better accuracies. This opens exciting possibilities of ML-guided analysis into nucleus independent chemical shifts complimenting the latest tight-binding model for PAHs\cite{kilymis2020efficient}. Although our 100k training set is an adequate representation of the QM9 dataset, adaptive sampling method employed in \RRef{Gerrard2020} might be useful when using smaller training sets. Given the locality of the shielding property, it may be helpful to employ different machines\cite{ML_TDDFTEnrico2015} trained on {\it sp}, {\it sp}$^2$ and {\it sp}$^3$ C---to account for systematic deviations in each groups.
 Finally, one can always improve the QM9-NMR dataset by estimating the effects from geometries obtained at $\omega$B97XD with triple-zeta quality basis set. 

\section{Acknowledgments}
The authors thank 
Vipin Agarwal, 
Kaustubh R. Mote and
O. Anatole von Lilienfeld
for fruitful discussions. 
This project was funded by intramural funds at TIFR Hyderabad from 
the Department of Atomic Energy (DAE).
All calculations have been performed using the Helios computer cluster, 
which is an integral part of the MolDis Big Data facility, 
TIFR Hyderabad (https://moldis.tifrh.res.in/).

\section{Data Availability}
The data that support the findings of this study are openly
available in the MolDis repository, 
{\tt http://moldis.tifrh.res.in:3000/QM9NMR}.
For further details see 
supplementary information.
\section*{References}
\bibliography{lit} 

\begin{thebibliography}{99}%
\makeatletter
\providecommand \@ifxundefined [1]{%
 \@ifx{#1\undefined}
}%
\providecommand \@ifnum [1]{%
 \ifnum #1\expandafter \@firstoftwo
 \else \expandafter \@secondoftwo
 \fi
}%
\providecommand \@ifx [1]{%
 \ifx #1\expandafter \@firstoftwo
 \else \expandafter \@secondoftwo
 \fi
}%
\providecommand \natexlab [1]{#1}%
\providecommand \enquote  [1]{``#1''}%
\providecommand \bibnamefont  [1]{#1}%
\providecommand \bibfnamefont [1]{#1}%
\providecommand \citenamefont [1]{#1}%
\providecommand \href@noop [0]{\@secondoftwo}%
\providecommand \href [0]{\begingroup \@sanitize@url \@href}%
\providecommand \@href[1]{\@@startlink{#1}\@@href}%
\providecommand \@@href[1]{\endgroup#1\@@endlink}%
\providecommand \@sanitize@url [0]{\catcode `\\12\catcode `\$12\catcode
  `\&12\catcode `\#12\catcode `\^12\catcode `\_12\catcode `\%12\relax}%
\providecommand \@@startlink[1]{}%
\providecommand \@@endlink[0]{}%
\providecommand \url  [0]{\begingroup\@sanitize@url \@url }%
\providecommand \@url [1]{\endgroup\@href {#1}{\urlprefix }}%
\providecommand \urlprefix  [0]{URL }%
\providecommand \Eprint [0]{\href }%
\providecommand \doibase [0]{http://dx.doi.org/}%
\providecommand \selectlanguage [0]{\@gobble}%
\providecommand \bibinfo  [0]{\@secondoftwo}%
\providecommand \bibfield  [0]{\@secondoftwo}%
\providecommand \translation [1]{[#1]}%
\providecommand \BibitemOpen [0]{}%
\providecommand \bibitemStop [0]{}%
\providecommand \bibitemNoStop [0]{.\EOS\space}%
\providecommand \EOS [0]{\spacefactor3000\relax}%
\providecommand \BibitemShut  [1]{\csname bibitem#1\endcsname}%
\let\auto@bib@innerbib\@empty
\bibitem [{\citenamefont {Helgaker}, \citenamefont {Jaszu{\'n}ski},\ and\
  \citenamefont {Ruud}(1999)}]{helgaker1999ab}%
  \BibitemOpen
  \bibfield  {author} {\bibinfo {author} {\bibfnamefont {T.}~\bibnamefont
  {Helgaker}}, \bibinfo {author} {\bibfnamefont {M.}~\bibnamefont
  {Jaszu{\'n}ski}}, \ and\ \bibinfo {author} {\bibfnamefont {K.}~\bibnamefont
  {Ruud}},\ }\href {https://pubs.acs.org/doi/pdfplus/10.1021/cr960017t}
  {\bibfield  {journal} {\bibinfo  {journal} {Chem. Rev.}\ }\textbf {\bibinfo
  {volume} {99}},\ \bibinfo {pages} {293} (\bibinfo {year} {1999})}\BibitemShut
  {NoStop}%
\bibitem [{\citenamefont {Mulder}\ and\ \citenamefont
  {Filatov}(2010)}]{mulder2010nmr}%
  \BibitemOpen
  \bibfield  {author} {\bibinfo {author} {\bibfnamefont {F.~A.}\ \bibnamefont
  {Mulder}}\ and\ \bibinfo {author} {\bibfnamefont {M.}~\bibnamefont
  {Filatov}},\ }\href
  {https://pubs.rsc.org/en/content/articlelanding/2010/cs/b811366c/unauth#!divAbstract}
  {\bibfield  {journal} {\bibinfo  {journal} {Chem. Soc. Rev.}\ }\textbf
  {\bibinfo {volume} {39}},\ \bibinfo {pages} {578} (\bibinfo {year}
  {2010})}\BibitemShut {NoStop}%
\bibitem [{\citenamefont {Bagno}, \citenamefont {Rastrelli},\ and\
  \citenamefont {Saielli}(2003)}]{bagno2003predicting}%
  \BibitemOpen
  \bibfield  {author} {\bibinfo {author} {\bibfnamefont {A.}~\bibnamefont
  {Bagno}}, \bibinfo {author} {\bibfnamefont {F.}~\bibnamefont {Rastrelli}}, \
  and\ \bibinfo {author} {\bibfnamefont {G.}~\bibnamefont {Saielli}},\ }\href
  {https://pubs.acs.org/doi/abs/10.1021/jp0353284} {\bibfield  {journal}
  {\bibinfo  {journal} {J. Phys. Chem. A}\ }\textbf {\bibinfo {volume} {107}},\
  \bibinfo {pages} {9964} (\bibinfo {year} {2003})}\BibitemShut {NoStop}%
\bibitem [{\citenamefont {Novotny}\ \emph {et~al.}(2016)\citenamefont
  {Novotny}, \citenamefont {Sojka}, \citenamefont {Komorovsky}, \citenamefont
  {Necas},\ and\ \citenamefont {Marek}}]{novotny2016interpreting}%
  \BibitemOpen
  \bibfield  {author} {\bibinfo {author} {\bibfnamefont {J.}~\bibnamefont
  {Novotny}}, \bibinfo {author} {\bibfnamefont {M.}~\bibnamefont {Sojka}},
  \bibinfo {author} {\bibfnamefont {S.}~\bibnamefont {Komorovsky}}, \bibinfo
  {author} {\bibfnamefont {M.}~\bibnamefont {Necas}}, \ and\ \bibinfo {author}
  {\bibfnamefont {R.}~\bibnamefont {Marek}},\ }\href
  {https://pubs.acs.org/doi/abs/10.1021/jacs.6b02749} {\bibfield  {journal}
  {\bibinfo  {journal} {J. Am. Chem. Soc.}\ }\textbf {\bibinfo {volume}
  {138}},\ \bibinfo {pages} {8432} (\bibinfo {year} {2016})}\BibitemShut
  {NoStop}%
\bibitem [{\citenamefont {Bifulco}, \citenamefont {Gomez-Paloma},\ and\
  \citenamefont {Riccio}(2003)}]{bifulco2003configurational}%
  \BibitemOpen
  \bibfield  {author} {\bibinfo {author} {\bibfnamefont {G.}~\bibnamefont
  {Bifulco}}, \bibinfo {author} {\bibfnamefont {L.}~\bibnamefont
  {Gomez-Paloma}}, \ and\ \bibinfo {author} {\bibfnamefont {R.}~\bibnamefont
  {Riccio}},\ }\href
  {https://www.sciencedirect.com/science/article/pii/S0040403903018100}
  {\bibfield  {journal} {\bibinfo  {journal} {Tetrahedron Lett.}\ }\textbf
  {\bibinfo {volume} {44}},\ \bibinfo {pages} {7137} (\bibinfo {year}
  {2003})}\BibitemShut {NoStop}%
\bibitem [{\citenamefont {Cimino}\ \emph {et~al.}(2004)\citenamefont {Cimino},
  \citenamefont {Gomez-Paloma}, \citenamefont {Duca}, \citenamefont {Riccio},\
  and\ \citenamefont {Bifulco}}]{cimino2004comparison}%
  \BibitemOpen
  \bibfield  {author} {\bibinfo {author} {\bibfnamefont {P.}~\bibnamefont
  {Cimino}}, \bibinfo {author} {\bibfnamefont {L.}~\bibnamefont
  {Gomez-Paloma}}, \bibinfo {author} {\bibfnamefont {D.}~\bibnamefont {Duca}},
  \bibinfo {author} {\bibfnamefont {R.}~\bibnamefont {Riccio}}, \ and\ \bibinfo
  {author} {\bibfnamefont {G.}~\bibnamefont {Bifulco}},\ }\href
  {https://onlinelibrary.wiley.com/doi/abs/10.1002/mrc.1410} {\bibfield
  {journal} {\bibinfo  {journal} {Magn. Reson. Chem.}\ }\textbf {\bibinfo
  {volume} {42}},\ \bibinfo {pages} {S26} (\bibinfo {year} {2004})}\BibitemShut
  {NoStop}%
\bibitem [{\citenamefont {Seymour}\ \emph {et~al.}(2016)\citenamefont
  {Seymour}, \citenamefont {Middlemiss}, \citenamefont {Halat}, \citenamefont
  {Trease}, \citenamefont {Pell},\ and\ \citenamefont
  {Grey}}]{seymour2016characterizing}%
  \BibitemOpen
  \bibfield  {author} {\bibinfo {author} {\bibfnamefont {I.~D.}\ \bibnamefont
  {Seymour}}, \bibinfo {author} {\bibfnamefont {D.~S.}\ \bibnamefont
  {Middlemiss}}, \bibinfo {author} {\bibfnamefont {D.~M.}\ \bibnamefont
  {Halat}}, \bibinfo {author} {\bibfnamefont {N.~M.}\ \bibnamefont {Trease}},
  \bibinfo {author} {\bibfnamefont {A.~J.}\ \bibnamefont {Pell}}, \ and\
  \bibinfo {author} {\bibfnamefont {C.~P.}\ \bibnamefont {Grey}},\ }\href
  {https://pubs.acs.org/doi/abs/10.1021/jacs.6b05747} {\bibfield  {journal}
  {\bibinfo  {journal} {J. Am. Chem. Soc.}\ }\textbf {\bibinfo {volume}
  {138}},\ \bibinfo {pages} {9405} (\bibinfo {year} {2016})}\BibitemShut
  {NoStop}%
\bibitem [{\citenamefont {Bamine}\ \emph {et~al.}(2017)\citenamefont {Bamine},
  \citenamefont {Boivin}, \citenamefont {Boucher}, \citenamefont {Messinger},
  \citenamefont {Salager}, \citenamefont {Deschamps}, \citenamefont
  {Masquelier}, \citenamefont {Croguennec}, \citenamefont {M{\'e}n{\'e}trier},\
  and\ \citenamefont {Carlier}}]{bamine2017understanding}%
  \BibitemOpen
  \bibfield  {author} {\bibinfo {author} {\bibfnamefont {T.}~\bibnamefont
  {Bamine}}, \bibinfo {author} {\bibfnamefont {E.}~\bibnamefont {Boivin}},
  \bibinfo {author} {\bibfnamefont {F.}~\bibnamefont {Boucher}}, \bibinfo
  {author} {\bibfnamefont {R.~J.}\ \bibnamefont {Messinger}}, \bibinfo {author}
  {\bibfnamefont {E.}~\bibnamefont {Salager}}, \bibinfo {author} {\bibfnamefont
  {M.}~\bibnamefont {Deschamps}}, \bibinfo {author} {\bibfnamefont
  {C.}~\bibnamefont {Masquelier}}, \bibinfo {author} {\bibfnamefont
  {L.}~\bibnamefont {Croguennec}}, \bibinfo {author} {\bibfnamefont
  {M.}~\bibnamefont {M{\'e}n{\'e}trier}}, \ and\ \bibinfo {author}
  {\bibfnamefont {D.}~\bibnamefont {Carlier}},\ }\href
  {https://pubs.acs.org/doi/abs/10.1021/acs.jpcc.6b11747} {\bibfield  {journal}
  {\bibinfo  {journal} {J. Phys. Chem. C}\ }\textbf {\bibinfo {volume} {121}},\
  \bibinfo {pages} {3219} (\bibinfo {year} {2017})}\BibitemShut {NoStop}%
\bibitem [{\citenamefont {Molchanov}\ \emph {et~al.}(2018)\citenamefont
  {Molchanov}, \citenamefont {Rowicki}, \citenamefont {Gryff-Keller},\ and\
  \citenamefont {Ko{\'z}mi{\'n}ski}}]{molchanov2018conformational}%
  \BibitemOpen
  \bibfield  {author} {\bibinfo {author} {\bibfnamefont {S.}~\bibnamefont
  {Molchanov}}, \bibinfo {author} {\bibfnamefont {T.}~\bibnamefont {Rowicki}},
  \bibinfo {author} {\bibfnamefont {A.}~\bibnamefont {Gryff-Keller}}, \ and\
  \bibinfo {author} {\bibfnamefont {W.}~\bibnamefont {Ko{\'z}mi{\'n}ski}},\
  }\href {https://pubs.acs.org/doi/abs/10.1021/acs.jpca.8b06722} {\bibfield
  {journal} {\bibinfo  {journal} {J. Phys. Chem. A}\ }\textbf {\bibinfo
  {volume} {122}},\ \bibinfo {pages} {7832} (\bibinfo {year}
  {2018})}\BibitemShut {NoStop}%
\bibitem [{\citenamefont {Guzzo}\ \emph {et~al.}(2018)\citenamefont {Guzzo},
  \citenamefont {Rezende}, \citenamefont {Kartnaller}, \citenamefont
  {Carneiro}, \citenamefont {Stoyanov},\ and\ \citenamefont
  {da~Costa}}]{guzzo2018experimental}%
  \BibitemOpen
  \bibfield  {author} {\bibinfo {author} {\bibfnamefont {R.~N.}\ \bibnamefont
  {Guzzo}}, \bibinfo {author} {\bibfnamefont {M.~J.~C.}\ \bibnamefont
  {Rezende}}, \bibinfo {author} {\bibfnamefont {V.}~\bibnamefont {Kartnaller}},
  \bibinfo {author} {\bibfnamefont {J.~W. d.~M.}\ \bibnamefont {Carneiro}},
  \bibinfo {author} {\bibfnamefont {S.~R.}\ \bibnamefont {Stoyanov}}, \ and\
  \bibinfo {author} {\bibfnamefont {L.~M.}\ \bibnamefont {da~Costa}},\ }\href
  {https://www.sciencedirect.com/science/article/pii/S0022286017316587}
  {\bibfield  {journal} {\bibinfo  {journal} {J. Mol. Struc.}\ }\textbf
  {\bibinfo {volume} {1157}},\ \bibinfo {pages} {97} (\bibinfo {year}
  {2018})}\BibitemShut {NoStop}%
\bibitem [{\citenamefont {Sarotti}(2018)}]{sarotti2018structural}%
  \BibitemOpen
  \bibfield  {author} {\bibinfo {author} {\bibfnamefont {A.~M.}\ \bibnamefont
  {Sarotti}},\ }\href
  {https://pubs.rsc.org/en/content/articlelanding/2017/ob/c7ob02916k/unauth#!divAbstract}
  {\bibfield  {journal} {\bibinfo  {journal} {Org. Biomol. Chem.}\ }\textbf
  {\bibinfo {volume} {16}},\ \bibinfo {pages} {944} (\bibinfo {year}
  {2018})}\BibitemShut {NoStop}%
\bibitem [{\citenamefont {Lodewyk}, \citenamefont {Siebert},\ and\
  \citenamefont {Tantillo}(2012)}]{lodewyk2012computational}%
  \BibitemOpen
  \bibfield  {author} {\bibinfo {author} {\bibfnamefont {M.~W.}\ \bibnamefont
  {Lodewyk}}, \bibinfo {author} {\bibfnamefont {M.~R.}\ \bibnamefont
  {Siebert}}, \ and\ \bibinfo {author} {\bibfnamefont {D.~J.}\ \bibnamefont
  {Tantillo}},\ }\href@noop {} {\bibfield  {journal} {\bibinfo  {journal}
  {Chem. Rev.}\ }\textbf {\bibinfo {volume} {112}},\ \bibinfo {pages} {1839}
  (\bibinfo {year} {2012})}\BibitemShut {NoStop}%
\bibitem [{\citenamefont {Grimme}\ \emph {et~al.}(2017)\citenamefont {Grimme},
  \citenamefont {Bannwarth}, \citenamefont {Dohm}, \citenamefont {Hansen},
  \citenamefont {Pisarek}, \citenamefont {Pracht}, \citenamefont {Seibert},\
  and\ \citenamefont {Neese}}]{grimme2017fully}%
  \BibitemOpen
  \bibfield  {author} {\bibinfo {author} {\bibfnamefont {S.}~\bibnamefont
  {Grimme}}, \bibinfo {author} {\bibfnamefont {C.}~\bibnamefont {Bannwarth}},
  \bibinfo {author} {\bibfnamefont {S.}~\bibnamefont {Dohm}}, \bibinfo {author}
  {\bibfnamefont {A.}~\bibnamefont {Hansen}}, \bibinfo {author} {\bibfnamefont
  {J.}~\bibnamefont {Pisarek}}, \bibinfo {author} {\bibfnamefont
  {P.}~\bibnamefont {Pracht}}, \bibinfo {author} {\bibfnamefont
  {J.}~\bibnamefont {Seibert}}, \ and\ \bibinfo {author} {\bibfnamefont
  {F.}~\bibnamefont {Neese}},\ }\href
  {https://onlinelibrary.wiley.com/doi/full/10.1002/anie.201708266} {\bibfield
  {journal} {\bibinfo  {journal} {Angew. Chem. Int. Ed.}\ }\textbf {\bibinfo
  {volume} {56}},\ \bibinfo {pages} {14763} (\bibinfo {year}
  {2017})}\BibitemShut {NoStop}%
\bibitem [{\citenamefont {Buevich}\ and\ \citenamefont
  {Elyashberg}(2018)}]{buevich2018towards}%
  \BibitemOpen
  \bibfield  {author} {\bibinfo {author} {\bibfnamefont {A.~V.}\ \bibnamefont
  {Buevich}}\ and\ \bibinfo {author} {\bibfnamefont {M.~E.}\ \bibnamefont
  {Elyashberg}},\ }\href
  {https://onlinelibrary.wiley.com/doi/full/10.1002/mrc.4645} {\bibfield
  {journal} {\bibinfo  {journal} {Magn. Reson. Chem.}\ }\textbf {\bibinfo
  {volume} {56}},\ \bibinfo {pages} {493} (\bibinfo {year} {2018})}\BibitemShut
  {NoStop}%
\bibitem [{\citenamefont {Lauro}\ \emph {et~al.}(2020)\citenamefont {Lauro},
  \citenamefont {Das}, \citenamefont {Riccio}, \citenamefont {Reddy},\ and\
  \citenamefont {Bifulco}}]{lauro2020dft}%
  \BibitemOpen
  \bibfield  {author} {\bibinfo {author} {\bibfnamefont {G.}~\bibnamefont
  {Lauro}}, \bibinfo {author} {\bibfnamefont {P.}~\bibnamefont {Das}}, \bibinfo
  {author} {\bibfnamefont {R.}~\bibnamefont {Riccio}}, \bibinfo {author}
  {\bibfnamefont {D.~S.}\ \bibnamefont {Reddy}}, \ and\ \bibinfo {author}
  {\bibfnamefont {G.}~\bibnamefont {Bifulco}},\ }\href
  {https://pubs.acs.org/doi/abs/10.1021/acs.joc.9b03129} {\bibfield  {journal}
  {\bibinfo  {journal} {J. Org. Chem.}\ }\textbf {\bibinfo {volume} {85}},\
  \bibinfo {pages} {3297} (\bibinfo {year} {2020})}\BibitemShut {NoStop}%
\bibitem [{\citenamefont {Keith}\ and\ \citenamefont
  {Bader}(1992)}]{Keith1992}%
  \BibitemOpen
  \bibfield  {author} {\bibinfo {author} {\bibfnamefont {T.~A.}\ \bibnamefont
  {Keith}}\ and\ \bibinfo {author} {\bibfnamefont {R.~F.}\ \bibnamefont
  {Bader}},\ }\href {\doibase 10.1016/0009-2614(92)85733-Q} {\bibfield
  {journal} {\bibinfo  {journal} {Chem. Phys. Lett.}\ }\textbf {\bibinfo
  {volume} {194}},\ \bibinfo {pages} {1} (\bibinfo {year} {1992})}\BibitemShut
  {NoStop}%
\bibitem [{\citenamefont {Keith}\ and\ \citenamefont
  {Bader}(1993)}]{Keith1993}%
  \BibitemOpen
  \bibfield  {author} {\bibinfo {author} {\bibfnamefont {T.~A.}\ \bibnamefont
  {Keith}}\ and\ \bibinfo {author} {\bibfnamefont {R.~F.}\ \bibnamefont
  {Bader}},\ }\href {\doibase 10.1016/0009-2614(93)89127-4} {\bibfield
  {journal} {\bibinfo  {journal} {Chem. Phys. Lett.}\ }\textbf {\bibinfo
  {volume} {210}},\ \bibinfo {pages} {223} (\bibinfo {year}
  {1993})}\BibitemShut {NoStop}%
\bibitem [{\citenamefont {Mauri}, \citenamefont {Pfrommer},\ and\ \citenamefont
  {Louie}(1996)}]{Mauri1996}%
  \BibitemOpen
  \bibfield  {author} {\bibinfo {author} {\bibfnamefont {F.}~\bibnamefont
  {Mauri}}, \bibinfo {author} {\bibfnamefont {B.~G.}\ \bibnamefont {Pfrommer}},
  \ and\ \bibinfo {author} {\bibfnamefont {S.~G.}\ \bibnamefont {Louie}},\
  }\href {\doibase 10.1103/PhysRevLett.77.5300} {\bibfield  {journal} {\bibinfo
   {journal} {Phys. Rev. Lett.}\ }\textbf {\bibinfo {volume} {77}},\ \bibinfo
  {pages} {5300} (\bibinfo {year} {1996})}\BibitemShut {NoStop}%
\bibitem [{\citenamefont {Gregor}, \citenamefont {Mauri},\ and\ \citenamefont
  {Car}(1999)}]{Gregor1999}%
  \BibitemOpen
  \bibfield  {author} {\bibinfo {author} {\bibfnamefont {T.}~\bibnamefont
  {Gregor}}, \bibinfo {author} {\bibfnamefont {F.}~\bibnamefont {Mauri}}, \
  and\ \bibinfo {author} {\bibfnamefont {R.}~\bibnamefont {Car}},\ }\href
  {\doibase 10.1063/1.479451} {\bibfield  {journal} {\bibinfo  {journal} {J.
  Chem. Phys.}\ }\textbf {\bibinfo {volume} {111}},\ \bibinfo {pages} {1815}
  (\bibinfo {year} {1999})}\BibitemShut {NoStop}%
\bibitem [{\citenamefont {Kutzelnigg}, \citenamefont {Fleischer},\ and\
  \citenamefont {Schindler}(1990)}]{kutzelnigg1990iglo}%
  \BibitemOpen
  \bibfield  {author} {\bibinfo {author} {\bibfnamefont {W.}~\bibnamefont
  {Kutzelnigg}}, \bibinfo {author} {\bibfnamefont {U.}~\bibnamefont
  {Fleischer}}, \ and\ \bibinfo {author} {\bibfnamefont {M.}~\bibnamefont
  {Schindler}},\ }in\ \href
  {https://link.springer.com/chapter/10.1007/978-3-642-75932-1_3} {\emph
  {\bibinfo {booktitle} {Deuterium and shift calculation}}}\ (\bibinfo
  {publisher} {Springer},\ \bibinfo {year} {1990})\ pp.\ \bibinfo {pages}
  {165--262}\BibitemShut {NoStop}%
\bibitem [{\citenamefont {Ditchfield}(1972)}]{Ditchfield1972}%
  \BibitemOpen
  \bibfield  {author} {\bibinfo {author} {\bibfnamefont {R.}~\bibnamefont
  {Ditchfield}},\ }\href {\doibase 10.1063/1.1677088} {\bibfield  {journal}
  {\bibinfo  {journal} {J. Chem. Phys.}\ }\textbf {\bibinfo {volume} {56}},\
  \bibinfo {pages} {5688} (\bibinfo {year} {1972})}\BibitemShut {NoStop}%
\bibitem [{\citenamefont {Hinchliffe}(1987)}]{hinchliffe1987ab}%
  \BibitemOpen
  \bibfield  {author} {\bibinfo {author} {\bibfnamefont {A.}~\bibnamefont
  {Hinchliffe}},\ }\href@noop {} {\emph {\bibinfo {title} {Ab initio
  Determination of Molecular Properties}}}\ (\bibinfo  {publisher} {A.
  Hilger},\ \bibinfo {year} {1987})\BibitemShut {NoStop}%
\bibitem [{\citenamefont {Gauss}(2000)}]{gaussmolecular}%
  \BibitemOpen
  \bibfield  {author} {\bibinfo {author} {\bibfnamefont {J.}~\bibnamefont
  {Gauss}},\ }\href
  {http://citeseerx.ist.psu.edu/viewdoc/download?doi=10.1.1.141.1291&rep=rep1&type=pdf}
  {\bibfield  {journal} {\bibinfo  {journal} {Modern Methods and Algorithms of
  Quantum Chemistry}\ }\textbf {\bibinfo {volume} {3}},\ \bibinfo {pages} {541}
  (\bibinfo {year} {2000})}\BibitemShut {NoStop}%
\bibitem [{\citenamefont {Mehring}(2012)}]{mehring2012high}%
  \BibitemOpen
  \bibfield  {author} {\bibinfo {author} {\bibfnamefont {M.}~\bibnamefont
  {Mehring}},\ }\href {https://books.google.co.in/books?id=BCrrCAAAQBAJ} {\emph
  {\bibinfo {title} {High Resolution NMR Spectroscopy in Solids}}},\ NMR Basic
  Principles and Progress\ (\bibinfo  {publisher} {Springer Berlin
  Heidelberg},\ \bibinfo {year} {2012})\BibitemShut {NoStop}%
\bibitem [{\citenamefont {Price}\ and\ \citenamefont
  {Stanton}(2002)}]{price2002computational}%
  \BibitemOpen
  \bibfield  {author} {\bibinfo {author} {\bibfnamefont {D.~R.}\ \bibnamefont
  {Price}}\ and\ \bibinfo {author} {\bibfnamefont {J.~F.}\ \bibnamefont
  {Stanton}},\ }\href {https://pubs.acs.org/doi/abs/10.1021/ol0200450}
  {\bibfield  {journal} {\bibinfo  {journal} {Org. Lett.}\ }\textbf {\bibinfo
  {volume} {4}},\ \bibinfo {pages} {2809} (\bibinfo {year} {2002})}\BibitemShut
  {NoStop}%
\bibitem [{\citenamefont {Flaig}\ \emph {et~al.}(2014)\citenamefont {Flaig},
  \citenamefont {Maurer}, \citenamefont {Hanni}, \citenamefont {Braunger},
  \citenamefont {Kick}, \citenamefont {Thubauville},\ and\ \citenamefont
  {Ochsenfeld}}]{Flaig2014}%
  \BibitemOpen
  \bibfield  {author} {\bibinfo {author} {\bibfnamefont {D.}~\bibnamefont
  {Flaig}}, \bibinfo {author} {\bibfnamefont {M.}~\bibnamefont {Maurer}},
  \bibinfo {author} {\bibfnamefont {M.}~\bibnamefont {Hanni}}, \bibinfo
  {author} {\bibfnamefont {K.}~\bibnamefont {Braunger}}, \bibinfo {author}
  {\bibfnamefont {L.}~\bibnamefont {Kick}}, \bibinfo {author} {\bibfnamefont
  {M.}~\bibnamefont {Thubauville}}, \ and\ \bibinfo {author} {\bibfnamefont
  {C.}~\bibnamefont {Ochsenfeld}},\ }\href
  {https://pubs.acs.org/doi/abs/10.1021/ct400780f} {\bibfield  {journal}
  {\bibinfo  {journal} {J. Chem. Theory Comput.}\ }\textbf {\bibinfo {volume}
  {10}},\ \bibinfo {pages} {572} (\bibinfo {year} {2014})}\BibitemShut
  {NoStop}%
\bibitem [{\citenamefont {Curtiss}, \citenamefont {Redfern},\ and\
  \citenamefont {Raghavachari}(2011)}]{curtiss2011gn}%
  \BibitemOpen
  \bibfield  {author} {\bibinfo {author} {\bibfnamefont {L.~A.}\ \bibnamefont
  {Curtiss}}, \bibinfo {author} {\bibfnamefont {P.~C.}\ \bibnamefont
  {Redfern}}, \ and\ \bibinfo {author} {\bibfnamefont {K.}~\bibnamefont
  {Raghavachari}},\ }\href
  {https://onlinelibrary.wiley.com/doi/full/10.1002/wcms.59} {\bibfield
  {journal} {\bibinfo  {journal} {Wiley Interdiscip. Rev. Comput. Mol. Sci.}\
  }\textbf {\bibinfo {volume} {1}},\ \bibinfo {pages} {810} (\bibinfo {year}
  {2011})}\BibitemShut {NoStop}%
\bibitem [{\citenamefont {Semenov}, \citenamefont {Samultsev},\ and\
  \citenamefont {Krivdin}(2019)}]{semenov2019calculation}%
  \BibitemOpen
  \bibfield  {author} {\bibinfo {author} {\bibfnamefont {V.~A.}\ \bibnamefont
  {Semenov}}, \bibinfo {author} {\bibfnamefont {D.~O.}\ \bibnamefont
  {Samultsev}}, \ and\ \bibinfo {author} {\bibfnamefont {L.~B.}\ \bibnamefont
  {Krivdin}},\ }\href {https://pubs.acs.org/doi/abs/10.1021/acs.jpca.9b06780}
  {\bibfield  {journal} {\bibinfo  {journal} {J. Phys. Chem. A}\ }\textbf
  {\bibinfo {volume} {123}},\ \bibinfo {pages} {8417} (\bibinfo {year}
  {2019})}\BibitemShut {NoStop}%
\bibitem [{\citenamefont {Wiitala}, \citenamefont {Hoye},\ and\ \citenamefont
  {Cramer}(2006)}]{wiitala2006hybrid}%
  \BibitemOpen
  \bibfield  {author} {\bibinfo {author} {\bibfnamefont {K.~W.}\ \bibnamefont
  {Wiitala}}, \bibinfo {author} {\bibfnamefont {T.~R.}\ \bibnamefont {Hoye}}, \
  and\ \bibinfo {author} {\bibfnamefont {C.~J.}\ \bibnamefont {Cramer}},\
  }\href {https://pubs.acs.org/doi/abs/10.1021/ct6001016} {\bibfield  {journal}
  {\bibinfo  {journal} {J. Chem. Theory Comput.}\ }\textbf {\bibinfo {volume}
  {2}},\ \bibinfo {pages} {1085} (\bibinfo {year} {2006})}\BibitemShut
  {NoStop}%
\bibitem [{\citenamefont {Adamo}\ and\ \citenamefont
  {Barone}(1998)}]{adamo1998exchange}%
  \BibitemOpen
  \bibfield  {author} {\bibinfo {author} {\bibfnamefont {C.}~\bibnamefont
  {Adamo}}\ and\ \bibinfo {author} {\bibfnamefont {V.}~\bibnamefont {Barone}},\
  }\href {https://aip.scitation.org/doi/abs/10.1063/1.475428} {\bibfield
  {journal} {\bibinfo  {journal} {J. Chem. Phys.}\ }\textbf {\bibinfo {volume}
  {108}},\ \bibinfo {pages} {664} (\bibinfo {year} {1998})}\BibitemShut
  {NoStop}%
\bibitem [{\citenamefont {Migda}\ and\ \citenamefont
  {Rys}(2004)}]{migda2004giao}%
  \BibitemOpen
  \bibfield  {author} {\bibinfo {author} {\bibfnamefont {W.}~\bibnamefont
  {Migda}}\ and\ \bibinfo {author} {\bibfnamefont {B.}~\bibnamefont {Rys}},\
  }\href {https://onlinelibrary.wiley.com/doi/abs/10.1002/mrc.1366} {\bibfield
  {journal} {\bibinfo  {journal} {Magn. Reson. Chem.}\ }\textbf {\bibinfo
  {volume} {42}},\ \bibinfo {pages} {459} (\bibinfo {year} {2004})}\BibitemShut
  {NoStop}%
\bibitem [{\citenamefont {V{\'a}zquez}(2002)}]{vazquez2002giao}%
  \BibitemOpen
  \bibfield  {author} {\bibinfo {author} {\bibfnamefont {S.}~\bibnamefont
  {V{\'a}zquez}},\ }\href
  {https://pubs.rsc.org/en/content/articlehtml/2002/p2/b207060j} {\bibfield
  {journal} {\bibinfo  {journal} {J. Chem. Soc., Perkin Trans. 2}\ ,\ \bibinfo
  {pages} {2100}} (\bibinfo {year} {2002})}\BibitemShut {NoStop}%
\bibitem [{\citenamefont {Wiberg}\ \emph {et~al.}(1999)\citenamefont {Wiberg},
  \citenamefont {Hammer}, \citenamefont {Zilm},\ and\ \citenamefont
  {Cheeseman}}]{wiberg1999nmr}%
  \BibitemOpen
  \bibfield  {author} {\bibinfo {author} {\bibfnamefont {K.~B.}\ \bibnamefont
  {Wiberg}}, \bibinfo {author} {\bibfnamefont {J.~D.}\ \bibnamefont {Hammer}},
  \bibinfo {author} {\bibfnamefont {K.~W.}\ \bibnamefont {Zilm}}, \ and\
  \bibinfo {author} {\bibfnamefont {J.~R.}\ \bibnamefont {Cheeseman}},\ }\href
  {https://pubs.acs.org/doi/abs/10.1021/jo990423n} {\bibfield  {journal}
  {\bibinfo  {journal} {J. Org. Chem.}\ }\textbf {\bibinfo {volume} {64}},\
  \bibinfo {pages} {6394} (\bibinfo {year} {1999})}\BibitemShut {NoStop}%
\bibitem [{\citenamefont {Wiberg}\ \emph {et~al.}(2004)\citenamefont {Wiberg},
  \citenamefont {Hammer}, \citenamefont {Zilm}, \citenamefont {Keith},
  \citenamefont {Cheeseman},\ and\ \citenamefont {Duchamp}}]{wiberg2004nmr}%
  \BibitemOpen
  \bibfield  {author} {\bibinfo {author} {\bibfnamefont {K.~B.}\ \bibnamefont
  {Wiberg}}, \bibinfo {author} {\bibfnamefont {J.~D.}\ \bibnamefont {Hammer}},
  \bibinfo {author} {\bibfnamefont {K.~W.}\ \bibnamefont {Zilm}}, \bibinfo
  {author} {\bibfnamefont {T.~A.}\ \bibnamefont {Keith}}, \bibinfo {author}
  {\bibfnamefont {J.~R.}\ \bibnamefont {Cheeseman}}, \ and\ \bibinfo {author}
  {\bibfnamefont {J.~C.}\ \bibnamefont {Duchamp}},\ }\href
  {https://pubs.acs.org/doi/abs/10.1021/jo030258i} {\bibfield  {journal}
  {\bibinfo  {journal} {J. Org. Chem.}\ }\textbf {\bibinfo {volume} {69}},\
  \bibinfo {pages} {1086} (\bibinfo {year} {2004})}\BibitemShut {NoStop}%
\bibitem [{\citenamefont {Bassarello}\ \emph {et~al.}(2003)\citenamefont
  {Bassarello}, \citenamefont {Cimino}, \citenamefont {Gomez-Paloma},
  \citenamefont {Riccio},\ and\ \citenamefont
  {Bifulco}}]{bassarello2003simulation}%
  \BibitemOpen
  \bibfield  {author} {\bibinfo {author} {\bibfnamefont {C.}~\bibnamefont
  {Bassarello}}, \bibinfo {author} {\bibfnamefont {P.}~\bibnamefont {Cimino}},
  \bibinfo {author} {\bibfnamefont {L.}~\bibnamefont {Gomez-Paloma}}, \bibinfo
  {author} {\bibfnamefont {R.}~\bibnamefont {Riccio}}, \ and\ \bibinfo {author}
  {\bibfnamefont {G.}~\bibnamefont {Bifulco}},\ }\href
  {https://www.sciencedirect.com/science/article/pii/S0040402003015801}
  {\bibfield  {journal} {\bibinfo  {journal} {Tetrahedron}\ }\textbf {\bibinfo
  {volume} {59}},\ \bibinfo {pages} {9555} (\bibinfo {year}
  {2003})}\BibitemShut {NoStop}%
\bibitem [{\citenamefont {Sarotti}\ and\ \citenamefont
  {Pellegrinet}(2009)}]{sarotti2009multi}%
  \BibitemOpen
  \bibfield  {author} {\bibinfo {author} {\bibfnamefont {A.~M.}\ \bibnamefont
  {Sarotti}}\ and\ \bibinfo {author} {\bibfnamefont {S.~C.}\ \bibnamefont
  {Pellegrinet}},\ }\href {https://pubs.acs.org/doi/abs/10.1021/jo901234h}
  {\bibfield  {journal} {\bibinfo  {journal} {J. Org. Chem.}\ }\textbf
  {\bibinfo {volume} {74}},\ \bibinfo {pages} {7254} (\bibinfo {year}
  {2009})}\BibitemShut {NoStop}%
\bibitem [{\citenamefont {Sarotti}\ and\ \citenamefont
  {Pellegrinet}(2012)}]{sarotti2012application}%
  \BibitemOpen
  \bibfield  {author} {\bibinfo {author} {\bibfnamefont {A.~M.}\ \bibnamefont
  {Sarotti}}\ and\ \bibinfo {author} {\bibfnamefont {S.~C.}\ \bibnamefont
  {Pellegrinet}},\ }\href {https://pubs.acs.org/doi/abs/10.1021/jo3008447}
  {\bibfield  {journal} {\bibinfo  {journal} {J. Org. Chem.}\ }\textbf
  {\bibinfo {volume} {77}},\ \bibinfo {pages} {6059} (\bibinfo {year}
  {2012})}\BibitemShut {NoStop}%
\bibitem [{\citenamefont {Gerrard}\ \emph {et~al.}(2020)\citenamefont
  {Gerrard}, \citenamefont {Bratholm}, \citenamefont {Packer}, \citenamefont
  {Mulholland}, \citenamefont {Glowacki},\ and\ \citenamefont
  {Butts}}]{Gerrard2020}%
  \BibitemOpen
  \bibfield  {author} {\bibinfo {author} {\bibfnamefont {W.}~\bibnamefont
  {Gerrard}}, \bibinfo {author} {\bibfnamefont {L.~A.}\ \bibnamefont
  {Bratholm}}, \bibinfo {author} {\bibfnamefont {M.~J.}\ \bibnamefont
  {Packer}}, \bibinfo {author} {\bibfnamefont {A.~J.}\ \bibnamefont
  {Mulholland}}, \bibinfo {author} {\bibfnamefont {D.~R.}\ \bibnamefont
  {Glowacki}}, \ and\ \bibinfo {author} {\bibfnamefont {C.~P.}\ \bibnamefont
  {Butts}},\ }\href {\doibase 10.1039/C9SC03854J} {\bibfield  {journal}
  {\bibinfo  {journal} {Chem. Sci.}\ }\textbf {\bibinfo {volume} {11}},\
  \bibinfo {pages} {508} (\bibinfo {year} {2020})}\BibitemShut {NoStop}%
\bibitem [{\citenamefont {Cobas}(2020)}]{cobas2020nmr}%
  \BibitemOpen
  \bibfield  {author} {\bibinfo {author} {\bibfnamefont {C.}~\bibnamefont
  {Cobas}},\ }\href@noop {} {\bibfield  {journal} {\bibinfo  {journal} {Magn.
  Reson. Chem.}\ } (\bibinfo {year} {2020})}\BibitemShut {NoStop}%
\bibitem [{\citenamefont {Bratholm}\ \emph {et~al.}(2020)\citenamefont
  {Bratholm}, \citenamefont {Gerrard}, \citenamefont {Anderson}, \citenamefont
  {Bai}, \citenamefont {Choi}, \citenamefont {Dang}, \citenamefont {Hanchar},
  \citenamefont {Howard}, \citenamefont {Huard}, \citenamefont {Kim} \emph
  {et~al.}}]{bratholm2020community}%
  \BibitemOpen
  \bibfield  {author} {\bibinfo {author} {\bibfnamefont {L.~A.}\ \bibnamefont
  {Bratholm}}, \bibinfo {author} {\bibfnamefont {W.}~\bibnamefont {Gerrard}},
  \bibinfo {author} {\bibfnamefont {B.}~\bibnamefont {Anderson}}, \bibinfo
  {author} {\bibfnamefont {S.}~\bibnamefont {Bai}}, \bibinfo {author}
  {\bibfnamefont {S.}~\bibnamefont {Choi}}, \bibinfo {author} {\bibfnamefont
  {L.}~\bibnamefont {Dang}}, \bibinfo {author} {\bibfnamefont {P.}~\bibnamefont
  {Hanchar}}, \bibinfo {author} {\bibfnamefont {A.}~\bibnamefont {Howard}},
  \bibinfo {author} {\bibfnamefont {G.}~\bibnamefont {Huard}}, \bibinfo
  {author} {\bibfnamefont {S.}~\bibnamefont {Kim}},  \emph {et~al.},\ }\href
  {https://arxiv.org/abs/2008.05994} {\bibfield  {journal} {\bibinfo  {journal}
  {arXiv preprint arXiv:2008.05994}\ } (\bibinfo {year} {2020})}\BibitemShut
  {NoStop}%
\bibitem [{\citenamefont {Rupp}, \citenamefont {Ramakrishnan},\ and\
  \citenamefont {von Lilienfeld}(2015)}]{Rupp2015}%
  \BibitemOpen
  \bibfield  {author} {\bibinfo {author} {\bibfnamefont {M.}~\bibnamefont
  {Rupp}}, \bibinfo {author} {\bibfnamefont {R.}~\bibnamefont {Ramakrishnan}},
  \ and\ \bibinfo {author} {\bibfnamefont {O.~A.}\ \bibnamefont {von
  Lilienfeld}},\ }\href
  {https://pubs.acs.org/doi/abs/10.1021/acs.jpclett.5b01456} {\bibfield
  {journal} {\bibinfo  {journal} {J. Phys. Chem. Lett.}\ }\textbf {\bibinfo
  {volume} {6}},\ \bibinfo {pages} {3309} (\bibinfo {year} {2015})}\BibitemShut
  {NoStop}%
\bibitem [{\citenamefont {Rupp}\ \emph {et~al.}(2012)\citenamefont {Rupp},
  \citenamefont {Tkatchenko}, \citenamefont {M{\"u}ller},\ and\ \citenamefont
  {von Lilienfeld}}]{Rupp2012}%
  \BibitemOpen
  \bibfield  {author} {\bibinfo {author} {\bibfnamefont {M.}~\bibnamefont
  {Rupp}}, \bibinfo {author} {\bibfnamefont {A.}~\bibnamefont {Tkatchenko}},
  \bibinfo {author} {\bibfnamefont {K.-R.}\ \bibnamefont {M{\"u}ller}}, \ and\
  \bibinfo {author} {\bibfnamefont {O.~A.}\ \bibnamefont {von Lilienfeld}},\
  }\href {https://journals.aps.org/prl/abstract/10.1103/PhysRevLett.108.058301}
  {\bibfield  {journal} {\bibinfo  {journal} {Phys. Rev. Lett.}\ }\textbf
  {\bibinfo {volume} {108}},\ \bibinfo {pages} {058301} (\bibinfo {year}
  {2012})}\BibitemShut {NoStop}%
\bibitem [{\citenamefont {Gao}\ \emph {et~al.}(2020)\citenamefont {Gao},
  \citenamefont {Zhang}, \citenamefont {Peng}, \citenamefont {Zhang},\ and\
  \citenamefont {Glezakou}}]{gao2020general}%
  \BibitemOpen
  \bibfield  {author} {\bibinfo {author} {\bibfnamefont {P.}~\bibnamefont
  {Gao}}, \bibinfo {author} {\bibfnamefont {J.}~\bibnamefont {Zhang}}, \bibinfo
  {author} {\bibfnamefont {Q.}~\bibnamefont {Peng}}, \bibinfo {author}
  {\bibfnamefont {J.}~\bibnamefont {Zhang}}, \ and\ \bibinfo {author}
  {\bibfnamefont {V.-A.}\ \bibnamefont {Glezakou}},\ }\href
  {https://pubs.acs.org/doi/abs/10.1021/acs.jcim.0c00388} {\bibfield  {journal}
  {\bibinfo  {journal} {J. Chem. Inf. Model.}\ } (\bibinfo {year}
  {2020})}\BibitemShut {NoStop}%
\bibitem [{\citenamefont {Ghosh}\ \emph {et~al.}(2019)\citenamefont {Ghosh},
  \citenamefont {Stuke}, \citenamefont {Todorovi{\'c}}, \citenamefont
  {J{\o}rgensen}, \citenamefont {Schmidt}, \citenamefont {Vehtari},\ and\
  \citenamefont {Rinke}}]{ghosh2019deep}%
  \BibitemOpen
  \bibfield  {author} {\bibinfo {author} {\bibfnamefont {K.}~\bibnamefont
  {Ghosh}}, \bibinfo {author} {\bibfnamefont {A.}~\bibnamefont {Stuke}},
  \bibinfo {author} {\bibfnamefont {M.}~\bibnamefont {Todorovi{\'c}}}, \bibinfo
  {author} {\bibfnamefont {P.~B.}\ \bibnamefont {J{\o}rgensen}}, \bibinfo
  {author} {\bibfnamefont {M.~N.}\ \bibnamefont {Schmidt}}, \bibinfo {author}
  {\bibfnamefont {A.}~\bibnamefont {Vehtari}}, \ and\ \bibinfo {author}
  {\bibfnamefont {P.}~\bibnamefont {Rinke}},\ }\href@noop {} {\bibfield
  {journal} {\bibinfo  {journal} {Adv. Sci.}\ }\textbf {\bibinfo {volume}
  {6}},\ \bibinfo {pages} {1801367} (\bibinfo {year} {2019})}\BibitemShut
  {NoStop}%
\bibitem [{\citenamefont {Westermayr}\ and\ \citenamefont
  {Marquetand}(2020)}]{westermayr2020deep}%
  \BibitemOpen
  \bibfield  {author} {\bibinfo {author} {\bibfnamefont {J.}~\bibnamefont
  {Westermayr}}\ and\ \bibinfo {author} {\bibfnamefont {P.}~\bibnamefont
  {Marquetand}},\ }\href@noop {} {\bibfield  {journal} {\bibinfo  {journal} {J.
  Chem. Phys.}\ }\textbf {\bibinfo {volume} {153}},\ \bibinfo {pages} {154112}
  (\bibinfo {year} {2020})}\BibitemShut {NoStop}%
\bibitem [{\citenamefont {Rankine}, \citenamefont {Madkhali},\ and\
  \citenamefont {Penfold}(2020)}]{rankine2020deep}%
  \BibitemOpen
  \bibfield  {author} {\bibinfo {author} {\bibfnamefont {C.~D.}\ \bibnamefont
  {Rankine}}, \bibinfo {author} {\bibfnamefont {M.~M.}\ \bibnamefont
  {Madkhali}}, \ and\ \bibinfo {author} {\bibfnamefont {T.~J.}\ \bibnamefont
  {Penfold}},\ }\href@noop {} {\bibfield  {journal} {\bibinfo  {journal} {J.
  Phys. Chem. A}\ } (\bibinfo {year} {2020})}\BibitemShut {NoStop}%
\bibitem [{\citenamefont {Ramakrishnan}\ \emph
  {et~al.}(2015{\natexlab{a}})\citenamefont {Ramakrishnan}, \citenamefont
  {Hartmann}, \citenamefont {Tapavicza},\ and\ \citenamefont {von
  Lilienfeld}}]{ML_TDDFTEnrico2015}%
  \BibitemOpen
  \bibfield  {author} {\bibinfo {author} {\bibfnamefont {R.}~\bibnamefont
  {Ramakrishnan}}, \bibinfo {author} {\bibfnamefont {M.}~\bibnamefont
  {Hartmann}}, \bibinfo {author} {\bibfnamefont {E.}~\bibnamefont {Tapavicza}},
  \ and\ \bibinfo {author} {\bibfnamefont {O.~A.}\ \bibnamefont {von
  Lilienfeld}},\ }\href {https://aip.scitation.org/doi/full/10.1063/1.4928757}
  {\bibfield  {journal} {\bibinfo  {journal} {J. Chem. Phys.}\ }\textbf
  {\bibinfo {volume} {143}},\ \bibinfo {pages} {084111} (\bibinfo {year}
  {2015}{\natexlab{a}})}\BibitemShut {NoStop}%
\bibitem [{\citenamefont {Xue}, \citenamefont {Barbatti},\ and\ \citenamefont
  {Dral}(2020)}]{xue2020machine}%
  \BibitemOpen
  \bibfield  {author} {\bibinfo {author} {\bibfnamefont {B.-X.}\ \bibnamefont
  {Xue}}, \bibinfo {author} {\bibfnamefont {M.}~\bibnamefont {Barbatti}}, \
  and\ \bibinfo {author} {\bibfnamefont {P.~O.}\ \bibnamefont {Dral}},\
  }\href@noop {} {\bibfield  {journal} {\bibinfo  {journal} {J. Phys. Chem. A}\
  }\textbf {\bibinfo {volume} {124}},\ \bibinfo {pages} {7199} (\bibinfo {year}
  {2020})}\BibitemShut {NoStop}%
\bibitem [{\citenamefont {Westermayr}\ \emph {et~al.}(2020)\citenamefont
  {Westermayr}, \citenamefont {Faber}, \citenamefont {Christensen},
  \citenamefont {von Lilienfeld},\ and\ \citenamefont
  {Marquetand}}]{westermayr2020neural}%
  \BibitemOpen
  \bibfield  {author} {\bibinfo {author} {\bibfnamefont {J.}~\bibnamefont
  {Westermayr}}, \bibinfo {author} {\bibfnamefont {F.~A.}\ \bibnamefont
  {Faber}}, \bibinfo {author} {\bibfnamefont {A.~S.}\ \bibnamefont
  {Christensen}}, \bibinfo {author} {\bibfnamefont {O.~A.}\ \bibnamefont {von
  Lilienfeld}}, \ and\ \bibinfo {author} {\bibfnamefont {P.}~\bibnamefont
  {Marquetand}},\ }\href@noop {} {\bibfield  {journal} {\bibinfo  {journal}
  {Mach. Learn.: Sci. and Technol.}\ }\textbf {\bibinfo {volume} {1}},\
  \bibinfo {pages} {025009} (\bibinfo {year} {2020})}\BibitemShut {NoStop}%
\bibitem [{\citenamefont {Pronobis}\ \emph {et~al.}(2018)\citenamefont
  {Pronobis}, \citenamefont {Sch{\"u}tt}, \citenamefont {Tkatchenko},\ and\
  \citenamefont {M{\"u}ller}}]{pronobis2018capturing}%
  \BibitemOpen
  \bibfield  {author} {\bibinfo {author} {\bibfnamefont {W.}~\bibnamefont
  {Pronobis}}, \bibinfo {author} {\bibfnamefont {K.~T.}\ \bibnamefont
  {Sch{\"u}tt}}, \bibinfo {author} {\bibfnamefont {A.}~\bibnamefont
  {Tkatchenko}}, \ and\ \bibinfo {author} {\bibfnamefont {K.-R.}\ \bibnamefont
  {M{\"u}ller}},\ }\href@noop {} {\bibfield  {journal} {\bibinfo  {journal}
  {Eur. Phys. J. B}\ }\textbf {\bibinfo {volume} {91}},\ \bibinfo {pages} {178}
  (\bibinfo {year} {2018})}\BibitemShut {NoStop}%
\bibitem [{\citenamefont {Huo}\ and\ \citenamefont
  {Rupp}(2017)}]{huo2017unified}%
  \BibitemOpen
  \bibfield  {author} {\bibinfo {author} {\bibfnamefont {H.}~\bibnamefont
  {Huo}}\ and\ \bibinfo {author} {\bibfnamefont {M.}~\bibnamefont {Rupp}},\
  }\href
  {https://www.researchgate.net/profile/Matthias_Rupp/publication/316429137_Unified_Representation_for_Machine_Learning_of_Molecules_and_Crystals/links/597ba6ff0f7e9b880293f6bf/Unified-Representation-for-Machine-Learning-of-Molecules-and-Crystals.pdf}
  {\bibfield  {journal} {\bibinfo  {journal} {arXiv preprint arXiv:1704.06439}\
  }\textbf {\bibinfo {volume} {13754}} (\bibinfo {year} {2017})}\BibitemShut
  {NoStop}%
\bibitem [{\citenamefont {Bart{\'o}k}\ \emph {et~al.}(2010)\citenamefont
  {Bart{\'o}k}, \citenamefont {Payne}, \citenamefont {Kondor},\ and\
  \citenamefont {Cs{\'a}nyi}}]{bartok2010gaussian}%
  \BibitemOpen
  \bibfield  {author} {\bibinfo {author} {\bibfnamefont {A.~P.}\ \bibnamefont
  {Bart{\'o}k}}, \bibinfo {author} {\bibfnamefont {M.~C.}\ \bibnamefont
  {Payne}}, \bibinfo {author} {\bibfnamefont {R.}~\bibnamefont {Kondor}}, \
  and\ \bibinfo {author} {\bibfnamefont {G.}~\bibnamefont {Cs{\'a}nyi}},\
  }\href {https://journals.aps.org/prl/abstract/10.1103/PhysRevLett.104.136403}
  {\bibfield  {journal} {\bibinfo  {journal} {Phys. Rev. Lett.}\ }\textbf
  {\bibinfo {volume} {104}},\ \bibinfo {pages} {136403} (\bibinfo {year}
  {2010})}\BibitemShut {NoStop}%
\bibitem [{\citenamefont {Bart{\'o}k}, \citenamefont {Kondor},\ and\
  \citenamefont {Cs{\'a}nyi}(2013)}]{Bartok2013}%
  \BibitemOpen
  \bibfield  {author} {\bibinfo {author} {\bibfnamefont {A.~P.}\ \bibnamefont
  {Bart{\'o}k}}, \bibinfo {author} {\bibfnamefont {R.}~\bibnamefont {Kondor}},
  \ and\ \bibinfo {author} {\bibfnamefont {G.}~\bibnamefont {Cs{\'a}nyi}},\
  }\href {https://journals.aps.org/prb/abstract/10.1103/PhysRevB.87.184115}
  {\bibfield  {journal} {\bibinfo  {journal} {Phys. Rev. B}\ }\textbf {\bibinfo
  {volume} {87}},\ \bibinfo {pages} {184115} (\bibinfo {year}
  {2013})}\BibitemShut {NoStop}%
\bibitem [{\citenamefont {De}\ \emph {et~al.}(2016)\citenamefont {De},
  \citenamefont {Bart{\'o}k}, \citenamefont {Cs{\'a}nyi},\ and\ \citenamefont
  {Ceriotti}}]{De2016}%
  \BibitemOpen
  \bibfield  {author} {\bibinfo {author} {\bibfnamefont {S.}~\bibnamefont
  {De}}, \bibinfo {author} {\bibfnamefont {A.~P.}\ \bibnamefont {Bart{\'o}k}},
  \bibinfo {author} {\bibfnamefont {G.}~\bibnamefont {Cs{\'a}nyi}}, \ and\
  \bibinfo {author} {\bibfnamefont {M.}~\bibnamefont {Ceriotti}},\ }\href
  {https://pubs.rsc.org/en/content/articlehtml/2016/cp/c6cp00415f} {\bibfield
  {journal} {\bibinfo  {journal} {Phys. Chem. Chem. Phys.}\ }\textbf {\bibinfo
  {volume} {18}},\ \bibinfo {pages} {13754} (\bibinfo {year}
  {2016})}\BibitemShut {NoStop}%
\bibitem [{\citenamefont {Paruzzo}\ \emph {et~al.}(2018)\citenamefont
  {Paruzzo}, \citenamefont {Hofstetter}, \citenamefont {Musil}, \citenamefont
  {De}, \citenamefont {Ceriotti},\ and\ \citenamefont {Emsley}}]{Paruzzo2018}%
  \BibitemOpen
  \bibfield  {author} {\bibinfo {author} {\bibfnamefont {F.~M.}\ \bibnamefont
  {Paruzzo}}, \bibinfo {author} {\bibfnamefont {A.}~\bibnamefont {Hofstetter}},
  \bibinfo {author} {\bibfnamefont {F.}~\bibnamefont {Musil}}, \bibinfo
  {author} {\bibfnamefont {S.}~\bibnamefont {De}}, \bibinfo {author}
  {\bibfnamefont {M.}~\bibnamefont {Ceriotti}}, \ and\ \bibinfo {author}
  {\bibfnamefont {L.}~\bibnamefont {Emsley}},\ }\href
  {https://www.nature.com/articles/s41467-018-06972-x} {\bibfield  {journal}
  {\bibinfo  {journal} {Nat. Commun.}\ }\textbf {\bibinfo {volume} {9}},\
  \bibinfo {pages} {1} (\bibinfo {year} {2018})}\BibitemShut {NoStop}%
\bibitem [{\citenamefont {Chaker}\ \emph {et~al.}(2019)\citenamefont {Chaker},
  \citenamefont {Salanne}, \citenamefont {Delaye},\ and\ \citenamefont
  {Charpentier}}]{chaker2019nmr}%
  \BibitemOpen
  \bibfield  {author} {\bibinfo {author} {\bibfnamefont {Z.}~\bibnamefont
  {Chaker}}, \bibinfo {author} {\bibfnamefont {M.}~\bibnamefont {Salanne}},
  \bibinfo {author} {\bibfnamefont {J.-M.}\ \bibnamefont {Delaye}}, \ and\
  \bibinfo {author} {\bibfnamefont {T.}~\bibnamefont {Charpentier}},\ }\href
  {https://pubs.rsc.org/ko/content/articlelanding/2019/cp/c9cp02803j/unauth#!divAbstract}
  {\bibfield  {journal} {\bibinfo  {journal} {Phys. Chem. Chem. Phys.}\
  }\textbf {\bibinfo {volume} {21}},\ \bibinfo {pages} {21709} (\bibinfo {year}
  {2019})}\BibitemShut {NoStop}%
\bibitem [{\citenamefont {Hartman}\ \emph {et~al.}(2016)\citenamefont
  {Hartman}, \citenamefont {Kudla}, \citenamefont {Day}, \citenamefont
  {Mueller},\ and\ \citenamefont {Beran}}]{hartman2016benchmark}%
  \BibitemOpen
  \bibfield  {author} {\bibinfo {author} {\bibfnamefont {J.~D.}\ \bibnamefont
  {Hartman}}, \bibinfo {author} {\bibfnamefont {R.~A.}\ \bibnamefont {Kudla}},
  \bibinfo {author} {\bibfnamefont {G.~M.}\ \bibnamefont {Day}}, \bibinfo
  {author} {\bibfnamefont {L.~J.}\ \bibnamefont {Mueller}}, \ and\ \bibinfo
  {author} {\bibfnamefont {G.~J.}\ \bibnamefont {Beran}},\ }\href
  {https://pubs.rsc.org/en/content/articlehtml/2016/cp/c6cp01831a} {\bibfield
  {journal} {\bibinfo  {journal} {Phys. Chem. Chem. Phys.}\ }\textbf {\bibinfo
  {volume} {18}},\ \bibinfo {pages} {21686} (\bibinfo {year}
  {2016})}\BibitemShut {NoStop}%
\bibitem [{\citenamefont {Faber}\ \emph {et~al.}(2018)\citenamefont {Faber},
  \citenamefont {Christensen}, \citenamefont {Huang},\ and\ \citenamefont {von
  Lilienfeld}}]{Faber2018}%
  \BibitemOpen
  \bibfield  {author} {\bibinfo {author} {\bibfnamefont {F.~A.}\ \bibnamefont
  {Faber}}, \bibinfo {author} {\bibfnamefont {A.~S.}\ \bibnamefont
  {Christensen}}, \bibinfo {author} {\bibfnamefont {B.}~\bibnamefont {Huang}},
  \ and\ \bibinfo {author} {\bibfnamefont {O.~A.}\ \bibnamefont {von
  Lilienfeld}},\ }\href {\doibase 10.1063/1.5020710} {\bibfield  {journal}
  {\bibinfo  {journal} {J. Chem. Phys.}\ }\textbf {\bibinfo {volume} {148}},\
  \bibinfo {pages} {241717} (\bibinfo {year} {2018})}\BibitemShut {NoStop}%
\bibitem [{\citenamefont {Ramakrishnan}\ \emph {et~al.}(2014)\citenamefont
  {Ramakrishnan}, \citenamefont {Dral}, \citenamefont {Rupp},\ and\
  \citenamefont {von Lilienfeld}}]{ramakrishnan2014quantum}%
  \BibitemOpen
  \bibfield  {author} {\bibinfo {author} {\bibfnamefont {R.}~\bibnamefont
  {Ramakrishnan}}, \bibinfo {author} {\bibfnamefont {P.~O.}\ \bibnamefont
  {Dral}}, \bibinfo {author} {\bibfnamefont {M.}~\bibnamefont {Rupp}}, \ and\
  \bibinfo {author} {\bibfnamefont {O.~A.}\ \bibnamefont {von Lilienfeld}},\
  }\href {https://www.nature.com/articles/sdata201422} {\bibfield  {journal}
  {\bibinfo  {journal} {Sci. Data}\ }\textbf {\bibinfo {volume} {1}},\ \bibinfo
  {pages} {140022} (\bibinfo {year} {2014})}\BibitemShut {NoStop}%
\bibitem [{\citenamefont {Dral}\ \emph {et~al.}(2020)\citenamefont {Dral},
  \citenamefont {Owens}, \citenamefont {Dral},\ and\ \citenamefont
  {Cs{\'a}nyi}}]{dral2020hierarchical}%
  \BibitemOpen
  \bibfield  {author} {\bibinfo {author} {\bibfnamefont {P.~O.}\ \bibnamefont
  {Dral}}, \bibinfo {author} {\bibfnamefont {A.}~\bibnamefont {Owens}},
  \bibinfo {author} {\bibfnamefont {A.}~\bibnamefont {Dral}}, \ and\ \bibinfo
  {author} {\bibfnamefont {G.}~\bibnamefont {Cs{\'a}nyi}},\ }\href@noop {}
  {\bibfield  {journal} {\bibinfo  {journal} {J. Chem. Phys.}\ }\textbf
  {\bibinfo {volume} {152}},\ \bibinfo {pages} {204110} (\bibinfo {year}
  {2020})}\BibitemShut {NoStop}%
\bibitem [{\citenamefont {Ramakrishnan}\ \emph
  {et~al.}(2015{\natexlab{b}})\citenamefont {Ramakrishnan}, \citenamefont
  {Dral}, \citenamefont {Rupp},\ and\ \citenamefont {von
  Lilienfeld}}]{ramakrishnan2015big}%
  \BibitemOpen
  \bibfield  {author} {\bibinfo {author} {\bibfnamefont {R.}~\bibnamefont
  {Ramakrishnan}}, \bibinfo {author} {\bibfnamefont {P.~O.}\ \bibnamefont
  {Dral}}, \bibinfo {author} {\bibfnamefont {M.}~\bibnamefont {Rupp}}, \ and\
  \bibinfo {author} {\bibfnamefont {O.~A.}\ \bibnamefont {von Lilienfeld}},\
  }\href {https://pubs.acs.org/doi/abs/10.1021/acs.jctc.5b00099} {\bibfield
  {journal} {\bibinfo  {journal} {J. Chem. Theory Comput.}\ }\textbf {\bibinfo
  {volume} {11}},\ \bibinfo {pages} {2087} (\bibinfo {year}
  {2015}{\natexlab{b}})}\BibitemShut {NoStop}%
\bibitem [{\citenamefont {Ruddigkeit}\ \emph {et~al.}(2012)\citenamefont
  {Ruddigkeit}, \citenamefont {Van~Deursen}, \citenamefont {Blum},\ and\
  \citenamefont {Reymond}}]{ruddigkeit2012enumeration}%
  \BibitemOpen
  \bibfield  {author} {\bibinfo {author} {\bibfnamefont {L.}~\bibnamefont
  {Ruddigkeit}}, \bibinfo {author} {\bibfnamefont {R.}~\bibnamefont
  {Van~Deursen}}, \bibinfo {author} {\bibfnamefont {L.~C.}\ \bibnamefont
  {Blum}}, \ and\ \bibinfo {author} {\bibfnamefont {J.-L.}\ \bibnamefont
  {Reymond}},\ }\href {https://pubs.acs.org/doi/abs/10.1021/ci300415d}
  {\bibfield  {journal} {\bibinfo  {journal} {J. Chem. Inf. Model.}\ }\textbf
  {\bibinfo {volume} {52}},\ \bibinfo {pages} {2864} (\bibinfo {year}
  {2012})}\BibitemShut {NoStop}%
\bibitem [{\citenamefont {Faber}\ \emph {et~al.}(2017)\citenamefont {Faber},
  \citenamefont {Hutchison}, \citenamefont {Huang}, \citenamefont {Gilmer},
  \citenamefont {Schoenholz}, \citenamefont {Dahl}, \citenamefont {Vinyals},
  \citenamefont {Kearnes}, \citenamefont {Riley},\ and\ \citenamefont {von
  Lilienfeld}}]{faber2017prediction}%
  \BibitemOpen
  \bibfield  {author} {\bibinfo {author} {\bibfnamefont {F.~A.}\ \bibnamefont
  {Faber}}, \bibinfo {author} {\bibfnamefont {L.}~\bibnamefont {Hutchison}},
  \bibinfo {author} {\bibfnamefont {B.}~\bibnamefont {Huang}}, \bibinfo
  {author} {\bibfnamefont {J.}~\bibnamefont {Gilmer}}, \bibinfo {author}
  {\bibfnamefont {S.~S.}\ \bibnamefont {Schoenholz}}, \bibinfo {author}
  {\bibfnamefont {G.~E.}\ \bibnamefont {Dahl}}, \bibinfo {author}
  {\bibfnamefont {O.}~\bibnamefont {Vinyals}}, \bibinfo {author} {\bibfnamefont
  {S.}~\bibnamefont {Kearnes}}, \bibinfo {author} {\bibfnamefont {P.~F.}\
  \bibnamefont {Riley}}, \ and\ \bibinfo {author} {\bibfnamefont {O.~A.}\
  \bibnamefont {von Lilienfeld}},\ }\href
  {https://pubs.acs.org/doi/abs/10.1021/acs.jctc.7b00577} {\bibfield  {journal}
  {\bibinfo  {journal} {J. Chem. Theory Comput.}\ }\textbf {\bibinfo {volume}
  {13}},\ \bibinfo {pages} {5255} (\bibinfo {year} {2017})}\BibitemShut
  {NoStop}%
\bibitem [{\citenamefont {Ramakrishnan}\ and\ \citenamefont {von
  Lilienfeld}(2017)}]{ramakrishnan2017machine}%
  \BibitemOpen
  \bibfield  {author} {\bibinfo {author} {\bibfnamefont {R.}~\bibnamefont
  {Ramakrishnan}}\ and\ \bibinfo {author} {\bibfnamefont {O.~A.}\ \bibnamefont
  {von Lilienfeld}},\ }\href
  {https://onlinelibrary.wiley.com/doi/abs/10.1002/9781119356059#page=248}
  {\bibfield  {journal} {\bibinfo  {journal} {Rev. Comput. Chem.}\ }\textbf
  {\bibinfo {volume} {30}},\ \bibinfo {pages} {225} (\bibinfo {year}
  {2017})}\BibitemShut {NoStop}%
\bibitem [{\citenamefont {Ramakrishnan}\ and\ \citenamefont {von
  Lilienfeld}(2015)}]{ramakrishnan2015many}%
  \BibitemOpen
  \bibfield  {author} {\bibinfo {author} {\bibfnamefont {R.}~\bibnamefont
  {Ramakrishnan}}\ and\ \bibinfo {author} {\bibfnamefont {O.~A.}\ \bibnamefont
  {von Lilienfeld}},\ }\href
  {https://www.ingentaconnect.com/content/scs/chimia/2015/00000069/00000004/art00005}
  {\bibfield  {journal} {\bibinfo  {journal} {CHIMIA}\ }\textbf {\bibinfo
  {volume} {69}},\ \bibinfo {pages} {182} (\bibinfo {year} {2015})}\BibitemShut
  {NoStop}%
\bibitem [{\citenamefont {Mauri}, \citenamefont {Consonni},\ and\ \citenamefont
  {Todeschini}(2017)}]{Mauri2017}%
  \BibitemOpen
  \bibfield  {author} {\bibinfo {author} {\bibfnamefont {A.}~\bibnamefont
  {Mauri}}, \bibinfo {author} {\bibfnamefont {V.}~\bibnamefont {Consonni}}, \
  and\ \bibinfo {author} {\bibfnamefont {R.}~\bibnamefont {Todeschini}},\
  }\enquote {\bibinfo {title} {Molecular descriptors},}\ in\ \href {\doibase
  10.1007/978-3-319-27282-5_51} {\emph {\bibinfo {booktitle} {Handbook of
  Computational Chemistry}}},\ \bibinfo {editor} {edited by\ \bibinfo {editor}
  {\bibfnamefont {J.}~\bibnamefont {Leszczynski}}, \bibinfo {editor}
  {\bibfnamefont {A.}~\bibnamefont {Kaczmarek-Kedziera}}, \bibinfo {editor}
  {\bibfnamefont {T.}~\bibnamefont {Puzyn}}, \bibinfo {editor} {\bibfnamefont
  {M.}~\bibnamefont {G.~Papadopoulos}}, \bibinfo {editor} {\bibfnamefont
  {H.}~\bibnamefont {Reis}}, \ and\ \bibinfo {editor} {\bibfnamefont
  {M.}~\bibnamefont {K.~Shukla}}}\ (\bibinfo  {publisher} {Springer
  International Publishing},\ \bibinfo {address} {Cham},\ \bibinfo {year}
  {2017})\ pp.\ \bibinfo {pages} {2065--2093}\BibitemShut {NoStop}%
\bibitem [{\citenamefont {Randi{\'c}}(1996)}]{randic1996molecular}%
  \BibitemOpen
  \bibfield  {author} {\bibinfo {author} {\bibfnamefont {M.}~\bibnamefont
  {Randi{\'c}}},\ }\href
  {https://link.springer.com/article/10.1007%2FBF01166727} {\bibfield
  {journal} {\bibinfo  {journal} {J. Math. Chem.}\ }\textbf {\bibinfo {volume}
  {19}},\ \bibinfo {pages} {375} (\bibinfo {year} {1996})}\BibitemShut
  {NoStop}%
\bibitem [{\citenamefont {Randi{\'c}}(1997)}]{randic1997characterization}%
  \BibitemOpen
  \bibfield  {author} {\bibinfo {author} {\bibfnamefont {M.}~\bibnamefont
  {Randi{\'c}}},\ }\href {https://pubs.acs.org/doi/abs/10.1021/ci960174t}
  {\bibfield  {journal} {\bibinfo  {journal} {J. Chem. Inform. Comput. Sci.}\
  }\textbf {\bibinfo {volume} {37}},\ \bibinfo {pages} {672} (\bibinfo {year}
  {1997})}\BibitemShut {NoStop}%
\bibitem [{\citenamefont {Pozdnyakov}\ \emph {et~al.}(2020)\citenamefont
  {Pozdnyakov}, \citenamefont {Willatt}, \citenamefont {Bart{\'o}k},
  \citenamefont {Ortner}, \citenamefont {Cs{\'a}nyi},\ and\ \citenamefont
  {Ceriotti}}]{pozdnyakov2020completeness}%
  \BibitemOpen
  \bibfield  {author} {\bibinfo {author} {\bibfnamefont {S.~N.}\ \bibnamefont
  {Pozdnyakov}}, \bibinfo {author} {\bibfnamefont {M.~J.}\ \bibnamefont
  {Willatt}}, \bibinfo {author} {\bibfnamefont {A.~P.}\ \bibnamefont
  {Bart{\'o}k}}, \bibinfo {author} {\bibfnamefont {C.}~\bibnamefont {Ortner}},
  \bibinfo {author} {\bibfnamefont {G.}~\bibnamefont {Cs{\'a}nyi}}, \ and\
  \bibinfo {author} {\bibfnamefont {M.}~\bibnamefont {Ceriotti}},\ }\href
  {https://arxiv.org/abs/2001.11696} {\bibfield  {journal} {\bibinfo  {journal}
  {arXiv preprint arXiv:2001.11696}\ } (\bibinfo {year} {2020})}\BibitemShut
  {NoStop}%
\bibitem [{\citenamefont {von Lilienfeld}\ \emph {et~al.}(2015)\citenamefont
  {von Lilienfeld}, \citenamefont {Ramakrishnan}, \citenamefont {Rupp},\ and\
  \citenamefont {Knoll}}]{von2015fourier}%
  \BibitemOpen
  \bibfield  {author} {\bibinfo {author} {\bibfnamefont {O.~A.}\ \bibnamefont
  {von Lilienfeld}}, \bibinfo {author} {\bibfnamefont {R.}~\bibnamefont
  {Ramakrishnan}}, \bibinfo {author} {\bibfnamefont {M.}~\bibnamefont {Rupp}},
  \ and\ \bibinfo {author} {\bibfnamefont {A.}~\bibnamefont {Knoll}},\ }\href
  {https://onlinelibrary.wiley.com/doi/full/10.1002/qua.24912} {\bibfield
  {journal} {\bibinfo  {journal} {Int. J. Quantum Chem.}\ }\textbf {\bibinfo
  {volume} {115}},\ \bibinfo {pages} {1084} (\bibinfo {year}
  {2015})}\BibitemShut {NoStop}%
\bibitem [{\citenamefont {Todeschini}\ and\ \citenamefont
  {Consonni}(2000)}]{Todeschini2000}%
  \BibitemOpen
  \bibfield  {author} {\bibinfo {author} {\bibfnamefont {R.}~\bibnamefont
  {Todeschini}}\ and\ \bibinfo {author} {\bibfnamefont {V.}~\bibnamefont
  {Consonni}},\ }\href {\doibase 10.1002/9783527613106} {\emph {\bibinfo
  {title} {{Handbook of Molecular Descriptors}}}},\ edited by\ \bibinfo
  {editor} {\bibfnamefont {R.}~\bibnamefont {Mannhold}}, \bibinfo {editor}
  {\bibfnamefont {H.}~\bibnamefont {Kubinyi}}, \ and\ \bibinfo {editor}
  {\bibfnamefont {H.}~\bibnamefont {Timmerman}},\ \bibinfo {series} {Methods
  and Principles in Medicinal Chemistry}, Vol.~\bibinfo {volume} {11}\
  (\bibinfo  {publisher} {Wiley-VCH},\ \bibinfo {address} {New York},\ \bibinfo
  {year} {2000})\ p.\ \bibinfo {pages} {688}\BibitemShut {NoStop}%
\bibitem [{\citenamefont {Behler}\ and\ \citenamefont
  {Parrinello}(2007)}]{behler2007generalized}%
  \BibitemOpen
  \bibfield  {author} {\bibinfo {author} {\bibfnamefont {J.}~\bibnamefont
  {Behler}}\ and\ \bibinfo {author} {\bibfnamefont {M.}~\bibnamefont
  {Parrinello}},\ }\href
  {https://journals.aps.org/prl/abstract/10.1103/PhysRevLett.98.146401}
  {\bibfield  {journal} {\bibinfo  {journal} {Phys. Rev. Lett.}\ }\textbf
  {\bibinfo {volume} {98}},\ \bibinfo {pages} {146401} (\bibinfo {year}
  {2007})}\BibitemShut {NoStop}%
\bibitem [{\citenamefont {Behler}(2011)}]{behler2011atom}%
  \BibitemOpen
  \bibfield  {author} {\bibinfo {author} {\bibfnamefont {J.}~\bibnamefont
  {Behler}},\ }\href {https://aip.scitation.org/doi/full/10.1063/1.3553717}
  {\bibfield  {journal} {\bibinfo  {journal} {J. Chem. Phys.}\ }\textbf
  {\bibinfo {volume} {134}},\ \bibinfo {pages} {074106} (\bibinfo {year}
  {2011})}\BibitemShut {NoStop}%
\bibitem [{\citenamefont {Engel}\ \emph {et~al.}(2019)\citenamefont {Engel},
  \citenamefont {Anelli}, \citenamefont {Hofstetter}, \citenamefont {Paruzzo},
  \citenamefont {Emsley},\ and\ \citenamefont {Ceriotti}}]{engel2019bayesian}%
  \BibitemOpen
  \bibfield  {author} {\bibinfo {author} {\bibfnamefont {E.~A.}\ \bibnamefont
  {Engel}}, \bibinfo {author} {\bibfnamefont {A.}~\bibnamefont {Anelli}},
  \bibinfo {author} {\bibfnamefont {A.}~\bibnamefont {Hofstetter}}, \bibinfo
  {author} {\bibfnamefont {F.}~\bibnamefont {Paruzzo}}, \bibinfo {author}
  {\bibfnamefont {L.}~\bibnamefont {Emsley}}, \ and\ \bibinfo {author}
  {\bibfnamefont {M.}~\bibnamefont {Ceriotti}},\ }\href
  {https://pubs.rsc.org/en/content/articlelanding/2019/cp/c9cp04489b/unauth#!divAbstract}
  {\bibfield  {journal} {\bibinfo  {journal} {Phys. Chem. Chem. Phys.}\
  }\textbf {\bibinfo {volume} {21}},\ \bibinfo {pages} {23385} (\bibinfo {year}
  {2019})}\BibitemShut {NoStop}%
\bibitem [{\citenamefont {Moussa}(2012)}]{Moussa2012}%
  \BibitemOpen
  \bibfield  {author} {\bibinfo {author} {\bibfnamefont {J.~E.}\ \bibnamefont
  {Moussa}},\ }\href
  {https://journals.aps.org/prl/abstract/10.1103/PhysRevLett.109.059801}
  {\bibfield  {journal} {\bibinfo  {journal} {Phys. Rev. Lett.}\ }\textbf
  {\bibinfo {volume} {109}},\ \bibinfo {pages} {059801} (\bibinfo {year}
  {2012})}\BibitemShut {NoStop}%
\bibitem [{\citenamefont {Hansen}\ \emph {et~al.}(2015)\citenamefont {Hansen},
  \citenamefont {Biegler}, \citenamefont {Ramakrishnan}, \citenamefont
  {Pronobis}, \citenamefont {von Lilienfeld}, \citenamefont {M{\"{u}}llerr},\
  and\ \citenamefont {Tkatchenko}}]{hansen2015machine}%
  \BibitemOpen
  \bibfield  {author} {\bibinfo {author} {\bibfnamefont {K.}~\bibnamefont
  {Hansen}}, \bibinfo {author} {\bibfnamefont {F.}~\bibnamefont {Biegler}},
  \bibinfo {author} {\bibfnamefont {R.}~\bibnamefont {Ramakrishnan}}, \bibinfo
  {author} {\bibfnamefont {W.}~\bibnamefont {Pronobis}}, \bibinfo {author}
  {\bibfnamefont {O.~A.}\ \bibnamefont {von Lilienfeld}}, \bibinfo {author}
  {\bibfnamefont {K.-R.}\ \bibnamefont {M{\"{u}}llerr}}, \ and\ \bibinfo
  {author} {\bibfnamefont {A.}~\bibnamefont {Tkatchenko}},\ }\href
  {https://pubs.acs.org/doi/abs/10.1021/acs.jpclett.5b00831} {\bibfield
  {journal} {\bibinfo  {journal} {J. Phys. Chem. Lett.}\ }\textbf {\bibinfo
  {volume} {6}},\ \bibinfo {pages} {2326} (\bibinfo {year} {2015})}\BibitemShut
  {NoStop}%
\bibitem [{\citenamefont {Huang}\ and\ \citenamefont {von
  Lilienfeld}(2016)}]{huang2016communication}%
  \BibitemOpen
  \bibfield  {author} {\bibinfo {author} {\bibfnamefont {B.}~\bibnamefont
  {Huang}}\ and\ \bibinfo {author} {\bibfnamefont {O.~A.}\ \bibnamefont {von
  Lilienfeld}},\ }\href {https://aip.scitation.org/doi/full/10.1063/1.4964627}
  {\bibfield  {journal} {\bibinfo  {journal} {J. Chem. Phys.}\ }\textbf
  {\bibinfo {volume} {145}},\ \bibinfo {pages} {161102} (\bibinfo {year}
  {2016})}\BibitemShut {NoStop}%
\bibitem [{\citenamefont {Pronobis}, \citenamefont {Tkatchenko},\ and\
  \citenamefont {M{\"{u}}llerr}(2018)}]{pronobis2018many}%
  \BibitemOpen
  \bibfield  {author} {\bibinfo {author} {\bibfnamefont {W.}~\bibnamefont
  {Pronobis}}, \bibinfo {author} {\bibfnamefont {A.}~\bibnamefont
  {Tkatchenko}}, \ and\ \bibinfo {author} {\bibfnamefont {K.-R.}\ \bibnamefont
  {M{\"{u}}llerr}},\ }\href
  {https://pubs.acs.org/doi/abs/10.1021/acs.jctc.8b00110} {\bibfield  {journal}
  {\bibinfo  {journal} {J. Chem. Theory Comput.}\ }\textbf {\bibinfo {volume}
  {14}},\ \bibinfo {pages} {2991} (\bibinfo {year} {2018})}\BibitemShut
  {NoStop}%
\bibitem [{\citenamefont {Ditchfield}(1974)}]{ditchfield1974self}%
  \BibitemOpen
  \bibfield  {author} {\bibinfo {author} {\bibfnamefont {R.}~\bibnamefont
  {Ditchfield}},\ }\href
  {https://www.tandfonline.com/doi/abs/10.1080/00268977400100711} {\bibfield
  {journal} {\bibinfo  {journal} {Mol. Phys.}\ }\textbf {\bibinfo {volume}
  {27}},\ \bibinfo {pages} {789} (\bibinfo {year} {1974})}\BibitemShut
  {NoStop}%
\bibitem [{\citenamefont {Wolinski}, \citenamefont {Hinton},\ and\
  \citenamefont {Pulay}(1990)}]{wolinski1990efficient}%
  \BibitemOpen
  \bibfield  {author} {\bibinfo {author} {\bibfnamefont {K.}~\bibnamefont
  {Wolinski}}, \bibinfo {author} {\bibfnamefont {J.~F.}\ \bibnamefont
  {Hinton}}, \ and\ \bibinfo {author} {\bibfnamefont {P.}~\bibnamefont
  {Pulay}},\ }\href {https://pubs.acs.org/doi/pdf/10.1021/ja00179a005}
  {\bibfield  {journal} {\bibinfo  {journal} {J. Am. Chem. Soc.}\ }\textbf
  {\bibinfo {volume} {112}},\ \bibinfo {pages} {8251} (\bibinfo {year}
  {1990})}\BibitemShut {NoStop}%
\bibitem [{\citenamefont {Cheeseman}\ \emph {et~al.}(1996)\citenamefont
  {Cheeseman}, \citenamefont {Trucks}, \citenamefont {Keith},\ and\
  \citenamefont {Frisch}}]{cheeseman1996comparison}%
  \BibitemOpen
  \bibfield  {author} {\bibinfo {author} {\bibfnamefont {J.~R.}\ \bibnamefont
  {Cheeseman}}, \bibinfo {author} {\bibfnamefont {G.~W.}\ \bibnamefont
  {Trucks}}, \bibinfo {author} {\bibfnamefont {T.~A.}\ \bibnamefont {Keith}}, \
  and\ \bibinfo {author} {\bibfnamefont {M.~J.}\ \bibnamefont {Frisch}},\
  }\href {https://aip.scitation.org/doi/abs/10.1063/1.471789} {\bibfield
  {journal} {\bibinfo  {journal} {J. Chem. Phys.}\ }\textbf {\bibinfo {volume}
  {104}},\ \bibinfo {pages} {5497} (\bibinfo {year} {1996})}\BibitemShut
  {NoStop}%
\bibitem [{\citenamefont {Frisch}\ \emph {et~al.}(2016)\citenamefont {Frisch},
  \citenamefont {Trucks}, \citenamefont {Schlegel}, \citenamefont {Scuseria},
  \citenamefont {Robb}, \citenamefont {Cheeseman}, \citenamefont {Scalmani},
  \citenamefont {Barone}, \citenamefont {Petersson}, \citenamefont {Nakatsuji}
  \emph {et~al.}}]{G16short}%
  \BibitemOpen
  \bibfield  {author} {\bibinfo {author} {\bibfnamefont {M.~J.}\ \bibnamefont
  {Frisch}}, \bibinfo {author} {\bibfnamefont {G.~W.}\ \bibnamefont {Trucks}},
  \bibinfo {author} {\bibfnamefont {H.~B.}\ \bibnamefont {Schlegel}}, \bibinfo
  {author} {\bibfnamefont {G.~E.}\ \bibnamefont {Scuseria}}, \bibinfo {author}
  {\bibfnamefont {M.~A.}\ \bibnamefont {Robb}}, \bibinfo {author}
  {\bibfnamefont {J.~R.}\ \bibnamefont {Cheeseman}}, \bibinfo {author}
  {\bibfnamefont {G.}~\bibnamefont {Scalmani}}, \bibinfo {author}
  {\bibfnamefont {V.}~\bibnamefont {Barone}}, \bibinfo {author} {\bibfnamefont
  {G.~A.}\ \bibnamefont {Petersson}}, \bibinfo {author} {\bibfnamefont
  {H.}~\bibnamefont {Nakatsuji}},  \emph {et~al.},\ }\href@noop {} {\bibfield
  {journal} {\bibinfo  {journal} {Revision A}\ }\textbf {\bibinfo {volume} {3}}
  (\bibinfo {year} {2016})}\BibitemShut {NoStop}%
\bibitem [{\citenamefont {Stewart}(2016)}]{MOPAC}%
  \BibitemOpen
  \bibfield  {author} {\bibinfo {author} {\bibfnamefont {J.~J.}\ \bibnamefont
  {Stewart}},\ }\href@noop {} {\enquote {\bibinfo {title} {Mopac2016},}\ }
  (\bibinfo {year} {2016}),\ \bibinfo {note} {stewart Computational Chemistry,
  Colorado Springs, CO, USA, HTTP://OpenMOPAC.net}\BibitemShut {NoStop}%
\bibitem [{\citenamefont {Tomasi}, \citenamefont {Mennucci},\ and\
  \citenamefont {Cammi}(2005)}]{tomasi2005quantum}%
  \BibitemOpen
  \bibfield  {author} {\bibinfo {author} {\bibfnamefont {J.}~\bibnamefont
  {Tomasi}}, \bibinfo {author} {\bibfnamefont {B.}~\bibnamefont {Mennucci}}, \
  and\ \bibinfo {author} {\bibfnamefont {R.}~\bibnamefont {Cammi}},\ }\href
  {https://pubs.acs.org/doi/full/10.1021/cr9904009} {\bibfield  {journal}
  {\bibinfo  {journal} {Chem. Rev.}\ }\textbf {\bibinfo {volume} {105}},\
  \bibinfo {pages} {2999} (\bibinfo {year} {2005})}\BibitemShut {NoStop}%
\bibitem [{\citenamefont {Chakraborty}, \citenamefont {Kayastha},\ and\
  \citenamefont {Ramakrishnan}(2019)}]{chakraborty2019chemical}%
  \BibitemOpen
  \bibfield  {author} {\bibinfo {author} {\bibfnamefont {S.}~\bibnamefont
  {Chakraborty}}, \bibinfo {author} {\bibfnamefont {P.}~\bibnamefont
  {Kayastha}}, \ and\ \bibinfo {author} {\bibfnamefont {R.}~\bibnamefont
  {Ramakrishnan}},\ }\href
  {https://aip.scitation.org/doi/abs/10.1063/1.5088083} {\bibfield  {journal}
  {\bibinfo  {journal} {J. Chem. Phys.}\ }\textbf {\bibinfo {volume} {150}},\
  \bibinfo {pages} {114106} (\bibinfo {year} {2019})}\BibitemShut {NoStop}%
\bibitem [{\citenamefont {Fink}\ and\ \citenamefont
  {Reymond}(2007)}]{fink2007virtual}%
  \BibitemOpen
  \bibfield  {author} {\bibinfo {author} {\bibfnamefont {T.}~\bibnamefont
  {Fink}}\ and\ \bibinfo {author} {\bibfnamefont {J.-L.}\ \bibnamefont
  {Reymond}},\ }\href {https://pubs.acs.org/doi/abs/10.1021/ci600423u}
  {\bibfield  {journal} {\bibinfo  {journal} {J. Chem. Inf. Model.}\ }\textbf
  {\bibinfo {volume} {47}},\ \bibinfo {pages} {342} (\bibinfo {year}
  {2007})}\BibitemShut {NoStop}%
\bibitem [{\citenamefont {Blum}\ and\ \citenamefont
  {Reymond}(2009)}]{blum2009970}%
  \BibitemOpen
  \bibfield  {author} {\bibinfo {author} {\bibfnamefont {L.~C.}\ \bibnamefont
  {Blum}}\ and\ \bibinfo {author} {\bibfnamefont {J.-L.}\ \bibnamefont
  {Reymond}},\ }\href {https://pubs.acs.org/doi/abs/10.1021/ja902302h}
  {\bibfield  {journal} {\bibinfo  {journal} {J. Am. Chem. Soc.}\ }\textbf
  {\bibinfo {volume} {131}},\ \bibinfo {pages} {8732} (\bibinfo {year}
  {2009})}\BibitemShut {NoStop}%
\bibitem [{\citenamefont {Blum}, \citenamefont {van Deursen},\ and\
  \citenamefont {Reymond}(2011)}]{blum2011visualisation}%
  \BibitemOpen
  \bibfield  {author} {\bibinfo {author} {\bibfnamefont {L.~C.}\ \bibnamefont
  {Blum}}, \bibinfo {author} {\bibfnamefont {R.}~\bibnamefont {van Deursen}}, \
  and\ \bibinfo {author} {\bibfnamefont {J.-L.}\ \bibnamefont {Reymond}},\
  }\href {https://link.springer.com/article/10.1007/s10822-011-9436-y}
  {\bibfield  {journal} {\bibinfo  {journal} {J. Comput. Aided Mol. Des.}\
  }\textbf {\bibinfo {volume} {25}},\ \bibinfo {pages} {637} (\bibinfo {year}
  {2011})}\BibitemShut {NoStop}%
\bibitem [{\citenamefont {Corey}, \citenamefont {Czak{\'o}},\ and\
  \citenamefont {K{\"u}rti}(2007)}]{corey2007molecules}%
  \BibitemOpen
  \bibfield  {author} {\bibinfo {author} {\bibfnamefont {E.~J.}\ \bibnamefont
  {Corey}}, \bibinfo {author} {\bibfnamefont {B.}~\bibnamefont {Czak{\'o}}}, \
  and\ \bibinfo {author} {\bibfnamefont {L.}~\bibnamefont {K{\"u}rti}},\
  }\href@noop {} {\emph {\bibinfo {title} {Molecules and Medicine}}}\ (\bibinfo
   {publisher} {John Wiley \& Sons},\ \bibinfo {year} {2007})\BibitemShut
  {NoStop}%
\bibitem [{\citenamefont {O'Boyle}\ \emph {et~al.}(2011)\citenamefont
  {O'Boyle}, \citenamefont {Banck}, \citenamefont {James}, \citenamefont
  {Morley}, \citenamefont {Vandermeersch},\ and\ \citenamefont
  {Hutchison}}]{o2011open}%
  \BibitemOpen
  \bibfield  {author} {\bibinfo {author} {\bibfnamefont {N.~M.}\ \bibnamefont
  {O'Boyle}}, \bibinfo {author} {\bibfnamefont {M.}~\bibnamefont {Banck}},
  \bibinfo {author} {\bibfnamefont {C.~A.}\ \bibnamefont {James}}, \bibinfo
  {author} {\bibfnamefont {C.}~\bibnamefont {Morley}}, \bibinfo {author}
  {\bibfnamefont {T.}~\bibnamefont {Vandermeersch}}, \ and\ \bibinfo {author}
  {\bibfnamefont {G.~R.}\ \bibnamefont {Hutchison}},\ }\href
  {https://link.springer.com/article/10.1186/1758-2946-3-33} {\bibfield
  {journal} {\bibinfo  {journal} {J. Cheminformatics}\ }\textbf {\bibinfo
  {volume} {3}},\ \bibinfo {pages} {33} (\bibinfo {year} {2011})}\BibitemShut
  {NoStop}%
\bibitem [{\citenamefont {Hanwell}\ \emph {et~al.}(2012)\citenamefont
  {Hanwell}, \citenamefont {Curtis}, \citenamefont {Lonie}, \citenamefont
  {Vandermeersch}, \citenamefont {Zurek},\ and\ \citenamefont
  {Hutchison}}]{hanwell2012avogadro}%
  \BibitemOpen
  \bibfield  {author} {\bibinfo {author} {\bibfnamefont {M.~D.}\ \bibnamefont
  {Hanwell}}, \bibinfo {author} {\bibfnamefont {D.~E.}\ \bibnamefont {Curtis}},
  \bibinfo {author} {\bibfnamefont {D.~C.}\ \bibnamefont {Lonie}}, \bibinfo
  {author} {\bibfnamefont {T.}~\bibnamefont {Vandermeersch}}, \bibinfo {author}
  {\bibfnamefont {E.}~\bibnamefont {Zurek}}, \ and\ \bibinfo {author}
  {\bibfnamefont {G.~R.}\ \bibnamefont {Hutchison}},\ }\href
  {https://link.springer.com/article/10.1186/1758-2946-4-17} {\bibfield
  {journal} {\bibinfo  {journal} {J. Cheminformatics}\ }\textbf {\bibinfo
  {volume} {4}},\ \bibinfo {pages} {17} (\bibinfo {year} {2012})}\BibitemShut
  {NoStop}%
\bibitem [{\citenamefont {Halgren}(1996)}]{halgren1996merck}%
  \BibitemOpen
  \bibfield  {author} {\bibinfo {author} {\bibfnamefont {T.~A.}\ \bibnamefont
  {Halgren}},\ }\href
  {https://onlinelibrary.wiley.com/doi/abs/10.1002/(SICI)1096-987X(199604)17:5/6<490::AID-JCC1>3.0.CO;2-P}
  {\bibfield  {journal} {\bibinfo  {journal} {J. Comput. Chem.}\ }\textbf
  {\bibinfo {volume} {17}},\ \bibinfo {pages} {490} (\bibinfo {year}
  {1996})}\BibitemShut {NoStop}%
\bibitem [{\citenamefont {Himanen}\ \emph {et~al.}(2020)\citenamefont
  {Himanen}, \citenamefont {J{\"a}ger}, \citenamefont {Morooka}, \citenamefont
  {Federici~Canova}, \citenamefont {Ranawat}, \citenamefont {Gao},
  \citenamefont {Rinke},\ and\ \citenamefont {Foster}}]{dscribe}%
  \BibitemOpen
  \bibfield  {author} {\bibinfo {author} {\bibfnamefont {L.}~\bibnamefont
  {Himanen}}, \bibinfo {author} {\bibfnamefont {M.~O.~J.}\ \bibnamefont
  {J{\"a}ger}}, \bibinfo {author} {\bibfnamefont {E.~V.}\ \bibnamefont
  {Morooka}}, \bibinfo {author} {\bibfnamefont {F.}~\bibnamefont
  {Federici~Canova}}, \bibinfo {author} {\bibfnamefont {Y.~S.}\ \bibnamefont
  {Ranawat}}, \bibinfo {author} {\bibfnamefont {D.~Z.}\ \bibnamefont {Gao}},
  \bibinfo {author} {\bibfnamefont {P.}~\bibnamefont {Rinke}}, \ and\ \bibinfo
  {author} {\bibfnamefont {A.~S.}\ \bibnamefont {Foster}},\ }\href {\doibase
  10.1016/j.cpc.2019.106949} {\bibfield  {journal} {\bibinfo  {journal}
  {Comput. Phys. Comm.}\ }\textbf {\bibinfo {volume} {247}},\ \bibinfo {pages}
  {106949} (\bibinfo {year} {2020})}\BibitemShut {NoStop}%
\bibitem [{\citenamefont {Christensen}\ \emph {et~al.}(2017)\citenamefont
  {Christensen}, \citenamefont {Faber}, \citenamefont {Huang}, \citenamefont
  {Bratholm}, \citenamefont {Tkatchenko}, \citenamefont {M{\"{u}}ller},\ and\
  \citenamefont {von Lilienfeld}}]{QML}%
  \BibitemOpen
  \bibfield  {author} {\bibinfo {author} {\bibfnamefont {A.~S.}\ \bibnamefont
  {Christensen}}, \bibinfo {author} {\bibfnamefont {F.~A.}\ \bibnamefont
  {Faber}}, \bibinfo {author} {\bibfnamefont {B.}~\bibnamefont {Huang}},
  \bibinfo {author} {\bibfnamefont {L.~A.}\ \bibnamefont {Bratholm}}, \bibinfo
  {author} {\bibfnamefont {A.}~\bibnamefont {Tkatchenko}}, \bibinfo {author}
  {\bibfnamefont {K.-R.}\ \bibnamefont {M{\"{u}}ller}}, \ and\ \bibinfo
  {author} {\bibfnamefont {O.~A.}\ \bibnamefont {von Lilienfeld}},\ }\href@noop
  {} {\enquote {\bibinfo {title} {{QML}: A python toolkit for quantum machine
  learning},}\ } (\bibinfo {year} {2017}),\ \bibinfo {note}
  {\url{https://github.com/qmlcode/qml}}\BibitemShut {NoStop}%
\bibitem [{\citenamefont {Blackford}\ \emph {et~al.}(1997)\citenamefont
  {Blackford}, \citenamefont {Choi}, \citenamefont {Cleary}, \citenamefont
  {D'Azevedo}, \citenamefont {Demmel}, \citenamefont {Dhillon}, \citenamefont
  {Dongarra}, \citenamefont {Hammarling}, \citenamefont {Henry}, \citenamefont
  {Petitet}, \citenamefont {Stanley}, \citenamefont {Walker},\ and\
  \citenamefont {Whaley}}]{slug}%
  \BibitemOpen
  \bibfield  {author} {\bibinfo {author} {\bibfnamefont {L.~S.}\ \bibnamefont
  {Blackford}}, \bibinfo {author} {\bibfnamefont {J.}~\bibnamefont {Choi}},
  \bibinfo {author} {\bibfnamefont {A.}~\bibnamefont {Cleary}}, \bibinfo
  {author} {\bibfnamefont {E.}~\bibnamefont {D'Azevedo}}, \bibinfo {author}
  {\bibfnamefont {J.}~\bibnamefont {Demmel}}, \bibinfo {author} {\bibfnamefont
  {I.}~\bibnamefont {Dhillon}}, \bibinfo {author} {\bibfnamefont
  {J.}~\bibnamefont {Dongarra}}, \bibinfo {author} {\bibfnamefont
  {S.}~\bibnamefont {Hammarling}}, \bibinfo {author} {\bibfnamefont
  {G.}~\bibnamefont {Henry}}, \bibinfo {author} {\bibfnamefont
  {A.}~\bibnamefont {Petitet}}, \bibinfo {author} {\bibfnamefont
  {K.}~\bibnamefont {Stanley}}, \bibinfo {author} {\bibfnamefont
  {D.}~\bibnamefont {Walker}}, \ and\ \bibinfo {author} {\bibfnamefont {R.~C.}\
  \bibnamefont {Whaley}},\ }\href@noop {} {\emph {\bibinfo {title} {{ScaLAPACK}
  Users' Guide}}}\ (\bibinfo  {publisher} {Society for Industrial and Applied
  Mathematics},\ \bibinfo {address} {Philadelphia, PA},\ \bibinfo {year}
  {1997})\BibitemShut {NoStop}%
\bibitem [{\citenamefont {Krishnan}\ \emph {et~al.}(2020)\citenamefont
  {Krishnan}, \citenamefont {Ghosh}, \citenamefont {Gupta}, \citenamefont
  {Kayastha}, \citenamefont {Senthil}, \citenamefont {Das}, \citenamefont
  {Kandpal}, \citenamefont {Chakraborty}, \citenamefont {Gupta},\ and\
  \citenamefont {Ramakrishnan}}]{moldis}%
  \BibitemOpen
  \bibfield  {author} {\bibinfo {author} {\bibfnamefont {S.}~\bibnamefont
  {Krishnan}}, \bibinfo {author} {\bibfnamefont {A.}~\bibnamefont {Ghosh}},
  \bibinfo {author} {\bibfnamefont {M.}~\bibnamefont {Gupta}}, \bibinfo
  {author} {\bibfnamefont {P.}~\bibnamefont {Kayastha}}, \bibinfo {author}
  {\bibfnamefont {S.}~\bibnamefont {Senthil}}, \bibinfo {author} {\bibfnamefont
  {S.~K.}\ \bibnamefont {Das}}, \bibinfo {author} {\bibfnamefont {S.~C.}\
  \bibnamefont {Kandpal}}, \bibinfo {author} {\bibfnamefont {S.}~\bibnamefont
  {Chakraborty}}, \bibinfo {author} {\bibfnamefont {A.}~\bibnamefont {Gupta}},
  \ and\ \bibinfo {author} {\bibfnamefont {R.}~\bibnamefont {Ramakrishnan}},\
  }\href@noop {} {\enquote {\bibinfo {title} {{MolDis}: A big data analytics
  platform for molecular discovery},}\ } (\bibinfo {year} {2020}),\ \bibinfo
  {note} {\url{https://moldis.tifrh.res.in/}}\BibitemShut {NoStop}%
\bibitem [{\citenamefont {Molchanov}\ and\ \citenamefont
  {Gryff-Keller}(2017)}]{molchanov2017solvation}%
  \BibitemOpen
  \bibfield  {author} {\bibinfo {author} {\bibfnamefont {S.}~\bibnamefont
  {Molchanov}}\ and\ \bibinfo {author} {\bibfnamefont {A.}~\bibnamefont
  {Gryff-Keller}},\ }\href
  {https://pubs.acs.org/doi/abs/10.1021/acs.jpca.7b10110} {\bibfield  {journal}
  {\bibinfo  {journal} {J. Phys. Chem. A}\ }\textbf {\bibinfo {volume} {121}},\
  \bibinfo {pages} {9645} (\bibinfo {year} {2017})}\BibitemShut {NoStop}%
\bibitem [{\citenamefont {Langer}, \citenamefont {Goe{\ss}mann},\ and\
  \citenamefont {Rupp}(2020)}]{langer2020representations}%
  \BibitemOpen
  \bibfield  {author} {\bibinfo {author} {\bibfnamefont {M.~F.}\ \bibnamefont
  {Langer}}, \bibinfo {author} {\bibfnamefont {A.}~\bibnamefont
  {Goe{\ss}mann}}, \ and\ \bibinfo {author} {\bibfnamefont {M.}~\bibnamefont
  {Rupp}},\ }\href {https://arxiv.org/abs/2003.12081} {\bibfield  {journal}
  {\bibinfo  {journal} {arXiv preprint arXiv:2003.12081}\ } (\bibinfo {year}
  {2020})}\BibitemShut {NoStop}%
\bibitem [{\citenamefont {Kilymis}\ \emph {et~al.}(2020)\citenamefont
  {Kilymis}, \citenamefont {Bart{\'o}k}, \citenamefont {Pickard}, \citenamefont
  {Forse},\ and\ \citenamefont {Merlet}}]{kilymis2020efficient}%
  \BibitemOpen
  \bibfield  {author} {\bibinfo {author} {\bibfnamefont {D.}~\bibnamefont
  {Kilymis}}, \bibinfo {author} {\bibfnamefont {A.~P.}\ \bibnamefont
  {Bart{\'o}k}}, \bibinfo {author} {\bibfnamefont {C.~J.}\ \bibnamefont
  {Pickard}}, \bibinfo {author} {\bibfnamefont {A.~C.}\ \bibnamefont {Forse}},
  \ and\ \bibinfo {author} {\bibfnamefont {C.}~\bibnamefont {Merlet}},\ }\href
  {https://pubs.rsc.org/en/content/articlelanding/2020/cp/d0cp01705a/unauth#!divAbstract}
  {\bibfield  {journal} {\bibinfo  {journal} {Phys. Chem. Chem. Phys.}\ }
  (\bibinfo {year} {2020})}\BibitemShut {NoStop}%
\end{thebibliography}%
\end{document}